\pdfoutput=1

\documentclass[journal]{IEEEtran}

\usepackage{amsmath}
\usepackage{amssymb}
\usepackage{bm}
\usepackage[latin1]{inputenc}
\usepackage{graphicx}
\usepackage{multirow}
\usepackage{mathtools}
\usepackage{cite}
\usepackage{dsfont}	
\usepackage[bottom]{footmisc}	
\usepackage{xifthen}
\usepackage[caption=false,font=footnotesize]{subfig}

\usepackage{algorithm,algpseudocode}
\algrenewcommand\algorithmicrequire{\textbf{Input:}}
\algrenewcommand\algorithmicensure{\textbf{Output:}}

\usepackage{threeparttable}	

\newcommand{\V}[1]{\bm{#1}}
\newcommand{\Set}[1]{\mathcal{#1}}
\newcommand{\NSet}{\mathbb{N}}
\newcommand{\RSet}{\mathbb{R}}
\newcommand{\norm}[1]{\left\Vert #1 \right\Vert}

\newcommand{\deq}{\triangleq}

\newcommand{\diag}{\mathrm{diag}}

\newcommand{\T}[1][]{	\ifthenelse{\isempty{#1}}
	{\scriptsize{\textsc{t}}}
	{\raisebox{#1 cm}[0 cm][0 cm]{\scriptsize{\textsc{t}}}}
}

\newcommand{\be}{\begin{equation}}
\newcommand{\ee}{\end{equation}}
\newcommand{\ist}{\hspace*{.3mm}}
\newcommand{\rmv}{\hspace*{-.3mm}}

\usepackage{accents}

\DeclareMathOperator*{\argmax}{arg\,max}

\usepackage{balance}

\mathtoolsset{showonlyrefs = true}

\allowdisplaybreaks

\begin{document}

\title{Fusion of Sensor Measurements and \\[-.5mm]
Target-Provided Information in Multitarget Tracking\vspace{1mm}}

\author{
    \IEEEauthorblockN{Domenico~Gaglione,
    Paolo~Braca,~\IEEEmembership{Senior Member,~IEEE},
    Giovanni~Soldi,
    Florian~Meyer,~\IEEEmembership{Member,~IEEE},
    Franz~Hlawatsch,~\IEEEmembership{Fellow,~IEEE},
    and Moe Z. Win,~\IEEEmembership{Fellow,~IEEE}}
    \thanks{
This work was supported in part by the NATO Allied Command Transformation (ACT) under the DKOE project, 
by the Austrian Science Fund (FWF) under
grant P 32055-N31,
by the Czech Science Foundation (GA\v{C}R) under grant 17-19638S,
by the Office of Naval Research under grant N00014-21-1-2267,
and by the Army Research Office through the MIT Institute for Soldier Nanotechnologies, under Contract W911NF-13-D-0001.
Parts of this paper were previously presented at FUSION 2018, Cambridge, UK, July 2018 and at IEEE ICASSP 2019, Brighton, UK, May, 2019.
D.\ Gaglione, P.\ Braca, and G.\ Soldi are with the NATO Centre for Maritime Research and Experimentation (CMRE), La~Spezia, Italy (e-mail: domenico.gaglione@cmre.nato.int, paolo.braca@cmre.nato.int, giovanni.soldi@cmre.nato.int). 
F.\ Meyer is with the
Scripps Institution of Oceanography and the Department of Electrical and Computer Engineering, University of California San Diego, La Jolla, CA, USA (e-mail: flmeyer@ucsd.edu).
F.\ Hlawatsch is with the Institute of Telecommunications, TU Wien, Vienna, Austria (e-mail: franz.hlawatsch@tuwien.ac.at).
M.\ Z.\ Win is with the Laboratory for Information and Decision Systems (LIDS), Massachusetts Institute of Technology (MIT), Cambridge, MA, USA (e-mail: moewin@mit.edu).}
\vspace{-0.45 cm}
}

\maketitle

\begin{abstract}
Tracking multiple time-varying states
based on heterogeneous observations is a key problem in many
applications.
Here, we develop a statistical model and algorithm for
tracking an unknown number
of targets based on the probabilistic fusion of
observations from two classes of data sources. The first class, referred to as target-independent perception systems (TIPSs), 
consists of sensors that periodically produce noisy measurements of targets without requiring target cooperation.
The second class, referred to as target-dependent reporting systems (TDRSs), relies on cooperative targets that report noisy measurements of their state and
their identity. We present a joint TIPS--TDRS observation model that accounts for
observation-origin uncertainty, missed detections, false alarms,
and asynchronicity.
We then establish a factor graph that represents this 
observation model 
along with a state evolution model 
including target identities. Finally, by executing the sum-product algorithm on that factor graph, we obtain a scalable multitarget
tracking algorithm with inherent TIPS--TDRS
fusion. The performance
of the proposed algorithm is
evaluated using simulated data as well as real data from a maritime surveillance experiment.
\vspace{-.5mm}
\end{abstract}


\IEEEpeerreviewmaketitle

\section{Introduction}

\IEEEPARstart{M}{ultitarget tracking (MTT)} consists in estimating the number and the time-varying states
of multiple moving objects (targets) \cite{Mah:B07,LigginsHL:B08,BarWilTia:B11,Koch:B13,mallick13,vo15}.
To obtain satisfactory 
performance,
it is often necessary to fuse information from multiple heterogeneous data sources. 
Indeed, heterogeneous data fusion for MTT is a key task for
many applications including surveillance, robotics, and remote sensing \cite{LigginsHL:B08,BarWilTia:B11,Koch:B13}.

\vspace{-1mm}

\subsection{TIPS and TDRS}
\label{sec:TIPS}

In this paper, we focus on the fusion of observations produced by two different classes of 
data sources,
with the
objective of improving the overall MTT capabilities. The first class
will be referred to as target-independent 
perception systems (TIPSs). A TIPS consists of one or multiple sensors that rely on an active and periodic interrogation of the environment to acquire
target-related measurements.
Due to this active interrogation, a TIPS can also acquire
measurements related to targets that do not cooperate in any form. Active interrogation is based on signals that are either
transmitted by the TIPS itself or are signals of opportunity transmitted by other
sources.
The sensors constituting a TIPS may be heterogeneous and with different sensing modalities, e.g., radars, optical cameras, and sonars \cite{petersen12}.
The TIPS measurements 
are extracted in a preprocessing step from a noisy received signal.
As a consequence, they
are themselves noisy. Furthermore, due to errors made in the preprocessing step, certain measurements may not originate from a 
target (\textit{false alarms}) and certain
targets may not generate measurements (\textit{missed detections}) \cite{mallick13,vo15}. Since the targets do not cooperate, there is a TIPS measurement-origin uncertainty, i.e., it is not clear if a given 
TIPS measurement was generated by a target, and by which target.

The second class of 
data sources will be referred to as target-dependent reporting systems (TDRSs).
A TDRS relies on information autonomously transmitted by
cooperative targets:
each cooperative target is equipped
with a transmitter, which is identified by a code or \textit{ID}, and transmits messages to the TDRS---called \textit{reports}---that include
the ID and a noisy measurement of the target state.
Reports are received asynchronously by the TDRS;
moreover, due to imperfect communication 
channels between
the cooperative targets and the
TDRS, a report may be lost, i.e., not received at all, or it may be received but contain a corrupted ID and/or measurement.
Because of such corrupted IDs, the association between cooperative targets and reports is uncertain; this is similar to the TIPS measurement-origin uncertainty.
However,
each report must
originate from a cooperative target, i.e., it cannot be a false alarm.

TIPSs and TDRSs arise in
many applications. For example, in the maritime domain, coastal/harbor radars are TIPS sensors and the automatic identification system (AIS) \cite{tetreault05} is a TDRS,
and in air traffic control, primary surveillance radars are TIPS sensors and the automatic dependent surveillance broadcast (ADS-B) system \cite{strohmeier14} is a TDRS.
TIPSs and TDRSs
are mostly used as stand-alone systems, and the information they provide is usually combined
by fusing the respective estimates of the target tracks \cite{besada00,jidong04,danu07,garcia10,
jeon13,
Kaz:J17-11}. 
In this paper, by contrast, we address
the estimation of 
the
target states directly from the heterogeneous TIPS-TDRS observations.
This approach, which is known as observation-level fusion, can be expected to yield better performance \cite{chen03}.
An MTT algorithm fusing the observations from a single radar sensor and the AIS has been proposed in \cite{habtemariam15}.
This algorithm uses a joint probabilistic data association 
technique to 
cope with the unknown associations between targets and radar measurements as well as between targets and IDs, 
along with a gating technique to reduce the number of admissible 
associations.
However, 
the radar and AIS observations are assumed to be based on synchronous clocks, and the radar measurement rate is assumed to be a multiple of the AIS report rate.
In practice, 
these assumptions are 
often not satisfied.

\vspace{-1mm}

\subsection{Contributions and Paper Organization}
\label{sec:contr}

In our previous work \cite{MeyBraWilHla:J17,MeyKroWilLauHlaBraWin:J18,SolMeyBraHla:J19}, we presented an MTT algorithm that fuses measurements from multiple TIPS sensors and
is able 
to
confirm the existence of an
unknown and time-varying number of targets
and track the target states in the presence of TIPS measurement-origin uncertainty, missed detections, and false alarms.
This algorithm was
derived by formulating the multisensor MTT problem in a Bayesian
framework, representing the factorization of the joint posterior distribution 
by a factor graph, and efficiently marginalizing the joint posterior distribution via the sum-prod\-uct algorithm (SPA) \cite{KscFreLoe:01,Loe:04}.
The SPA
exploits conditional statistical independences for a drastic reduction of complexity. This results in an excellent scalability of the MTT algorithm of 
\cite{MeyBraWilHla:J17,MeyKroWilLauHlaBraWin:J18} with respect to the number of targets, the number
of TIPS sensors, and the number of measurements per sensor.
We note that an alternative framework for
the development of MTT algorithms is
constituted by random finite sets (RFSs) \cite{Vo:J07,Wil:15,FerWilSveXia:J21}.
Similarities and differences of an SPA-based derivation of MTT methods relative to an RFS-based derivation are discussed 
in \cite{MeyKroWilLauHlaBraWin:J18}.

Here, we propose an SPA-based framework and algorithm for \emph{MTT with TIPS-TDRS fusion}. More specifically, we extend the MTT framework and algorithm of
\cite{MeyBraWilHla:J17} to 
incorporate reports provided by a TDRS. 
Our key contributions are as follows:

\begin{itemize}

\vspace{.5mm}

\item We establish a statistical model of MTT based on measurements provided by multiple TIPS sensors and reports provided by a TDRS.
The TDRS reports are asynchronous and include 
the IDs from cooperative targets.

\vspace{1mm}

\item We represent this statistical model by a factor graph and use the SPA to develop a scalable message passing algorithm for MTT with TIPS-TDRS fusion.

\vspace{1mm}

\item We demonstrate performance advantages of MTT with TIPS-TDRS fusion using simulated data, and validate the proposed algorithm on
real data from a maritime surveillance experiment.
						
\vspace{.5mm}

\end{itemize}

\noindent
Parts of this work were presented
in our conference publications 
\cite{gaglione18} and \cite{SoldiGMHBFW:C19}.
This paper differs from
those publications in that it extends the formulation
beyond the maritime (i.e., AIS) domain; it introduces
an improved modeling for the TDRS IDs; it presents detailed derivations of the joint posterior distribution;
it presents the SPA messages in a more complete and detailed manner;
and it validates the performance of the proposed MTT algorithm in an additional simulated scenario and in a real maritime scenario.

The remainder of the paper is organized as follows. The basic notation and nomenclature are described in the next subsection.
Section~\ref{sec:problem_formulation} introduces 
the TIPS-TDRS fusion problem and 
outlines the proposed 
approach. 
The system model and its statistical formulation are described in Section~\ref{sec:system_model_statistical_formulation}. 
In Section~\ref{sec:proposed_algorithm}, we derive the joint posterior distribution and the corresponding factor graph. 
The proposed message passing algorithm
is presented in Section~\ref{sec:BP_approach}. 
Section~\ref{sec:results_simulations} provides an extensive
evaluation of the proposed
algorithm using simulated data. 
Section~\ref{sec:results_realdata} presents an application using real data from a maritime surveillance experiment.

\vspace{-1mm}

\subsection{Notation and Nomenclature}
\label{sec:notation}

Vectors are denoted by boldface lower-case letters (e.g., $\V{a}$),
matrices
by boldface upper-case letters (e.g., $\V{A}$),
and sets by calligraphic letters (e.g., $\Set{A}$).
The transpose is
written as $(\cdot)^{\T}\!$.
The Euclidean norm of vector $\V{a}$ is denoted by $\norm{\V{a}}$.
For a two-dimensional (2D) vector $\V{a}$, $\angle\V{a}$ is the angle 
defined clockwise and 
such that $\angle\V{a} \!=\! 0$
for $\V{a} \rmv=\rmv [0 \,\, 1]^{\T}$.
We write $\diag(a_{1},\ldots,$\linebreak $a_{N})$ for an $N \! \times \rmv N$ diagonal matrix with diagonal entries $a_{1},\ldots,$\linebreak $a_{N}$,
${\bf I}_{N}$ for the $N \! \times \! N$ identity matrix, 
and
$\V{0}$ for a zero vector.
We denote by $\mathds{1}(a)$ the indicator function of the event $a \!=\rmv 0$, i.e., $\mathds{1}(a) \rmv=\! 1$ for $a \!=\rmv 0$ and $\mathds{1}(a) \rmv=\rmv 0$ otherwise.
Finally, we denote the probability mass function (pmf) of a discrete random variable or vector by $p(\cdot)$ and the probability density function (pdf) of a continuous random variable
or vector by $f(\cdot)$; the latter notation will also be used for
a mixed pdf/pmf of both continuous and discrete random variables or vectors.

We use the term \textit{observation} generically for any target-related data
provided by a TIPS
sensor or a TDRS. 
Furthermore, the terms \textit{measurement} and \textit{report} are used for an observation provided by a TIPS
sensor and by a TDRS, respectively.
Finally, \textit{cluster} designates a group of reports, and 
\textit{self-measurement} 
the 
information related to the state of a cooperative target
that is contained in a report or cluster.

\section{TIPS-TDRS Fusion}
\label{sec:problem_formulation}

We consider a TIPS with $S$ sensors indexed by $s \rmv \in \rmv \Set{S} \rmv \deq \{1, \linebreak \ldots, S \}$. All TIPS sensors provide measurements at times 
$t_{n} \! = \! n T$, $n \! \in \! \NSet$. (The extension to the case where only some TIPS sensors provide measurements at time $t_{n}$ and/or the measurement times $t_{n}$
are spaced nonuniformly is straightforward.) As mentioned in Section~\ref{sec:TIPS}, the TIPS measurements are affected by noise, false alarms, missed detections,
and measurement-origin uncertainty.

In addition, we consider a TDRS, which is indexed by $s \rmv=\rmv 0$.
Each report received by the TDRS originates from a cooperative target and arises at an arbitrary time, i.e., asynchronously with respect to the TIPS measurements times $t_{n}$.
The set of IDs is defined as $\Set{D} \deq \{ 1, \ldots, D \}$ and is known by the TDRS.
In view of the imperfect target-to-TDRS communication channel, we distinguish between the intrinsic ID
embedded in
the target's transmitter, to be referred to as \textit{transmitter ID (TID)},
and the ID contained in a report received by the TDRS, to be referred to as \textit{report ID (RID)}.
To elucidate this distinction, let us consider a cooperative target with TID $d \in \Set{D}$ that transmits a report to the TDRS.
If the transmission is successful and without errors, the RID of the received report is itself $d \in \Set{D}$.
However, a transmission error will cause the RID 
to be 
equal to $d' \notin \Set{D}$ or equal to $d' \in \Set{D}$ with $d' \neq d$.
Therefore, even though the target is cooperative, the target-report association
is uncertain.
In addition, again due to the imperfect communication channel, also the self-measurement
contained in the TDRS report may be corrupted.
If this is detected by the TDRS, the report is 
discarded since the sole RID
is not meaningful for tracking.

\begin{figure}[!t]

\centering

\includegraphics{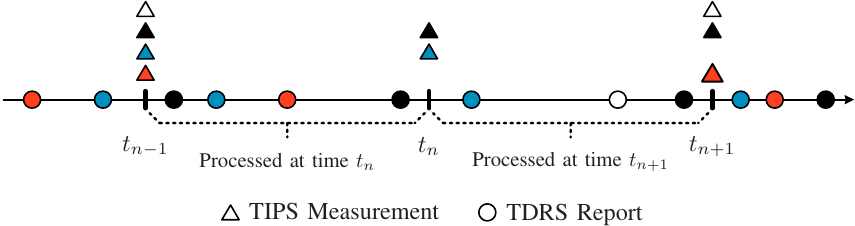}
\vspace{-.75mm}	

\caption{\hspace{-.7mm}Time diagram of TIPS measurements (from a single sensor) and TDRS reports during three consecutive time steps, assuming three cooperative targets located in the region of interest. 
The TIPS measurements and TDRS reports are represented by triangles and circles, respectively. The targets from which the TIPS measurements and TDRS reports originate 
are identified by different colors (black, blue, red).
White triangles represent clutter-generated TIPS measurements, and white circles 
represent TDRS reports with
RID $d \! \notin \! \Set{D}$.}
\label{fig:timeline}

\vspace{-2mm}

\end{figure}

Fig.~\ref{fig:timeline} shows an example of TIPS measurements
(from a single sensor) 
and TDRS reports that are available
over three consecutive time steps $t_{n - 1}, t_{n}, t_{n + 1}$. Note that at times $t_{n}$ and $t_{n + 1}$, one target is missed by the TIPS
sensor.
TDRS reports are received at arbitrary times, typically several of them---possibly also from the same cooperative target---between any two time steps.
The statistical model and estimation method 
that we will present allows the joint processing, at a considered time $t_{n}$, of all the TIPS measurements acquired at time $t_{n}$ and all the TDRS reports received during 
the time interval $( t_{n - 1},t_{n} ]$.

To enable this joint processing, we propose to group the reports
into clusters. More specifically, at time $t_{n}$, reports with an identical RID $d \! \in \! \Set{D}$ are grouped into the same
cluster, whereas reports with an RID $d \! \notin \! \Set{D}$ are not grouped.
If
there are no transmission errors,
then each cluster contains only reports generated by one corresponding cooperative target, and conversely, all reports transmitted by a given cooperative target are included in only one corresponding cluster.
Otherwise, some associations between reports and cooperative targets 
are incorrect. However, if the
RID errors are independent over time, 
incorrect associations tend to be resolved in future time steps.
As we will
show in Section~\ref{sec:first-scenario:comp}, this joint processing
can lead to
a more accurate estimation of the TIDs
compared to 
a sequential
processing in which the TDRS reports are not grouped and are 
processed as soon as they are received.

\section{System Model and Statistical Formulation}
\label{sec:system_model_statistical_formulation}

This section presents the system model underlying the proposed multisensor MTT algorithm and
a corresponding statistical (Bayesian) formulation.

\vspace{-1mm}

\subsection{Target States, Existence Indicators, and TIDs}
\label{sec:state}

We consider $K$ \textit{potential targets} (PTs) indexed by $k \! \in \! \Set{K} \deq \! \{1,\ldots,K \}$. Note that $K$ is the maximum possible number of actual 
targets.\footnote{Introducing 
a maximum possible number of targets $K$ leads to a predictable computational complexity. The resulting model is analogous to a multi-Bernoulli
birth process \cite{VoVoHoa:J17}. An alternative approach where the number of PTs is time-varying is presented in \cite{MeyKroWilLauHlaBraWin:J18}.}
The state of PT $k$ at time $n$ is represented by the vector $\V{x}_{n,k}$ and consists of the PT's position and, possibly, further
parameters, e.g., the PT's velocity.
The existence of PT $k$ at time $n$ is indicated by the binary variable $r_{n,k} \! \in \! \{0,1\}$, i.e.,
$r_{n,k} \! = \! 1$ if the PT exists and $r_{n,k} \! = \rmv 0$ otherwise. The state $\V{x}_{n,k}$ is formally defined also if 
$r_{n,k} \! = \rmv 0$. The time evolution of the state of a PT $k$ that existed at time $n \!-\! 1$ and still exists at time $n$ (i.e., for which $r_{n-1,k} \rmv=\rmv r_{n,k} \rmv=\! 1$) is modeled as
\begin{align}
	\V{x}_{n,k} = \V{\theta}(\V{x}_{n-1,k},\V{u}_{n,k}),
	\label{eq:dynamic_model1} 
\end{align}
where $\V{\theta}(\cdot)$ is some possibly nonlinear state transition function and $\V{u}_{n,k}$ is a driving process 
that is independent and identically distributed (iid) across $n$ and $k$ with a known pdf $f(\V{u})$.
This time evolution model (dynamic model)
defines the state transition
pdf $f (\V{x}_{n,k} | \V{x}_{n-1,k})$.

Each PT $k$ has a TID $\tau_{n,k}$. If PT $k$ is cooperative, then $\tau_{n,k} \! \in \! \Set{D} \! = \! \{ 1, \ldots, D \}$, 
and if it is noncooperative, we set
$\tau_{n,k} = 0$.
Nonexisting PTs (i.e., for which $r_{n,k} = 0$) are noncooperative.
Note that
an existing PT is noncooperative at time $n$ either if it is not equipped with a transmitter---thus, it is unable to send reports to the TDRS---or if it is provided 
with a transmitter but has not sent a report so far.
In the latter case,
an existing noncooperative PT can, at any time, become
cooperative by sending a report to the TDRS, and, consequently, its TID
can, at any time, transition from $\tau_{n-1,k} = 0$ to $\tau_{n,k} \in \Set{D}$.
The TID of a cooperative PT, instead, is time-invariant.
It will be convenient to combine the state
$\V{x}_{n,k}$, existence variable $r_{n,k}$, and TID $\tau_{n,k}$ of PT $k$ at time $n$ into the \emph{augmented state} 
$\V{y}_{n,k} \! \deq \! [\V{x}_{n,k}^{\T}, r_{n,k}, \tau_{n,k}]^{\T}\rmv$.
Finally, we define the vectors 
$\V{x}_{n} \rmv \deq \rmv [\V{x}_{n,1}^{\T}, \linebreak \ldots, \V{x}_{n,K}^{\T}]^{\T}\rmv$, 
$\V{r}_{n} \rmv \deq [r_{n,1},\ldots,r_{n,K}]^{\T}\rmv$,
$\V{\tau}_{n} \rmv \deq [\tau_{n,1}, \ldots, \tau_{n,K}]^{\T}\rmv$, and 
$\V{y}_{n} \rmv \deq \rmv [\V{y}_{n,1}^{\T}, \ldots,
\V{y}_{n,K}^{\T}]^{\T}$,
as well as $\V{x} \rmv \deq \rmv [\V{x}_{1}^{\T}, \ldots, \V{x}_{n}^{\T}]^{\T}\rmv$, $\V{r} \rmv \deq \rmv [\V{r}_{1}^{\T},\ldots,\V{r}_{n}^{\T}]^{\T}\rmv$,
$\V{\tau} \rmv \deq [\V{\tau}_{1}^{\T},\ldots,\V{\tau}_{n}^{\T}]^{\T}\rmv$, and $\V{y} \rmv \deq \rmv [\V{y}_{1}^{\T}, \ldots,\V{y}_{n}^{\T}]^{\T}$.

Under the assumption that the augmented states $\V{y}_{n}$ evolve according to a first-order Markov model \cite{BarWilTia:B11} and
the states, existence variables, and TIDs of different PTs evolve 
independently,\footnote{The 
independence assumption across PT states is 
commonly used in MTT algorithms \cite{BarWilTia:B11}. 
The independence across PT TIDs does not guarantee that each cooperative PT has a different estimated TID value $d \in \Set{D}$; 
in practice, however, this is ensured by the probabilistic data association algorithm described in Section \ref{subsec:target_observation_association}.}
the joint pdf of all the augmented states at all times factors 
as
\begin{align}
	\nonumber \\[-8mm]
	f(\V{y}) &= f(\V{y}_{0}) \rmv\prod_{n' = 1}^{n} \! f(\V{y}_{n'} | \V{y}_{n' - 1} ) \nonumber \\[-1mm]
	&= \prod_{k = 1}^{K} f(\V{y}_{0,k}) \rmv\prod_{n' = 1}^{n} \! f(\V{y}_{n',k} | \V{y}_{n' - 1,k} ),
	\label{eq:joint_pdf_augmented_state} \\[-5mm]
	\nonumber
\end{align}
where $f (\V{y}_{0}) = \prod_{k = 1}^{K} f (\V{y}_{0,k})$ is the prior joint pdf of all the augmented states at time $n \rmv=\rmv 0$.
To establish an expression of the transition pdf $f(\V{y}_{n,k} | \V{y}_{n-1,k})$, we assume that 
(i) conditioned on the previous
TID $\tau_{n-1,k}$, the previous existence variable $r_{n-1,k}$, and the current existence variable $r_{n,k}$, the current
TID $\tau_{n,k}$  
is statistically independent of the previous state $\V{x}_{n-1,k}$ and the current state $\V{x}_{n,k}$; 
and (ii) conditioned on
$\V{x}_{n-1,k}$ and 
$r_{n-1,k}$,
$\V{x}_{n,k}$ and
$r_{n,k}$ 
are statistically independent of 
$\tau_{n-1,k}$.
We then obtain
\begin{align}
	\nonumber \\[-6mm]
	f(\V{y}_{n,k} | \V{y}_{n-1,k}) &= f(\V{x}_{n,k},r_{n,k},\tau_{n,k} | \V{x}_{n-1,k},r_{n-1,k},\tau_{n-1,k}) \nonumber \\[.5mm]
	&= p(\tau_{n,k} |\V{x}_{n,k}, r_{n,k}, \V{x}_{n-1,k},r_{n-1,k},\tau_{n-1,k}) \nonumber \\[0mm]
	&\hspace{10mm} \times f(\V{x}_{n,k}, r_{n,k} | \V{x}_{n-1,k},r_{n-1,k}, \tau_{n-1,k}) \nonumber \\[.5mm]
	&= p(\tau_{n,k} |  r_{n,k}, r_{n-1,k}, \tau_{n-1,k}) \nonumber \\[0mm]
	&\hspace{10mm} \times f(\V{x}_{n,k}, r_{n,k} | \V{x}_{n-1,k}, r_{n-1,k}).
	\label{eq:augmented_state_transition_pdf} 
\end{align}

Next, we establish expressions of the factors in the product \eqref{eq:augmented_state_transition_pdf}.
To obtain an expression of $p (\tau_{n,k} | r_{n,k},r_{n-1,k},\tau_{n-1,k})$, we
recall that a nonexisting PT is noncooperative, 
i.e.,
\be
	p (\tau_{n,k} | r_{n,k} \!=\rmv  0, r_{n-1,k},\tau_{n-1,k}) = 
	\begin{cases}
		1 \ist, &\!\! \tau_{n,k} \!=\rmv 0 \ist, \\[0mm]
		0 \ist, &\!\! \tau_{n,k} \!\in\rmv \Set{D} \ist.
	\end{cases}
	\label{eq:single_target_label_pdf_1}
\ee
A PT that existed at time $n \!-\! 1$ and still exists at time $n$, and that was noncooperative at time $n \!-\! 1$, is, at time $n$, still noncooperative with probability 
$1 \!-\rmv p_{n,k}^{\text{t}}$ or cooperative with probability $p_{n,k}^{\text{t}}$; in the latter case, its TID can take on any value in $\Set{D}$ with equal probability, 
\vspace{-2mm}
i.e.,
\be
	p (\tau_{n,k} | r_{n,k} \!=\! 1,r_{n-1,k} \!=\! 1,\tau_{n-1,k} \!=\rmv 0) = 
	\begin{cases}
		1 \!-\rmv p_{n,k}^{\text{t}} \ist, &\!\!\!\rmv \tau_{n,k} \!=\rmv 0 \ist, \\[.5mm]
		p_{n,k}^{\text{t}} /\rmv D \ist, &\!\!\!\rmv \tau_{n,k} \!\in\! \Set{D} .
	\end{cases}
\vspace{.5mm}
\ee
Therefore, $p_{n,k}^{\text{t}}$ is the probability that the TID of 
the existing noncooperative PT $k$ transitions from $\tau_{n-1,k} = 0$ 
to any $\tau_{n,k} \in \Set{D}$,
or, in other words, the probability that the existing noncooperative PT $k$
becomes cooperative at time $n$.
Moreover,
a PT that existed at time $n \!-\! 1$ and still exists at time $n$, and that was cooperative at time $n \!-\! 1$, is still cooperative at time $n$, 
and its TID $\tau_{n-1,k} \in \Set{D}$ does not change, i.e.,
\be
	p (\tau_{n,k} | r_{n,k} \!=\! 1,r_{n-1,k} \!=\! 1,\tau_{n-1,k} \!=\rmv d \rmv\in\rmv \Set{D}) = 
	\begin{cases}
		1 \ist, &\!\!\! \tau_{n,k} \!=\rmv d \ist, \\[.5mm]
		0 \ist, &\!\!\! \text{otherwise} \ist.
	\end{cases}\nonumber  \nonumber
\ee
Finally, we assume that a newly
confirmed PT (i.e., for which $r_{n-1,k} = 0$ and $r_{n,k} = 1$) is
noncooperative with probability $p_{n,k}^{0}$ 
or
cooperative with probability $1 - p_{n,k}^{0}$; in the latter case, 
its TID can take on any value in $\Set{D}$ with equal probability, 
i.e.,
\begin{align}
	&p (\tau_{n,k} | r_{n,k} \!=\! 1,r_{n-1,k} \!=\rmv 0,\tau_{n-1,k}) \nonumber \\[.5mm]
	&\hspace{9mm}= \begin{cases}
		p_{n,k}^{0} \ist, &\!\! \tau_{n,k} \!=\rmv 0 \ist, \\[.5mm]
		\big( 1 \!-\rmv p_{n,k}^{0} \big) /\rmv D \ist, &\!\! \tau_{n,k} \!\in\! \Set{D}.
	\end{cases}
	\label{eq:single_target_label_pdf_2}
\end{align}

An expression of the second factor in \eqref{eq:augmented_state_transition_pdf}, $f (\V{x}_{n,k}, r_{n,k} | \linebreak \V{x}_{n-1,k}, r_{n-1,k})$, is provided by \cite[Eqs. (5) and (6)]{MeyBraWilHla:J17}.
These equations involve a survival probability $p_{n,k}^{\text{s}}$, a birth probability $p_{n,k}^{\text{b}}$, and a birth pdf $f_{\text{b}}(\V{x}_{n,k})$.

\vspace{-1mm}

\subsection{Observation Model}
\label{subsec:observation_model_vectors}

Let $\Set{S}_{0} \rmv \deq \Set{S} \cup \{0\} \rmv=\rmv \lbrace 0,1,\ldots,S \rbrace$ denote the index set of both the TIPS sensors 
($s \rmv\in\rmv \Set{S} \rmv=\rmv \lbrace 1,2,\ldots,S \rbrace$) and the TDRS ($s \rmv=\rmv 0$), which will be generically referred to as \textit{data sources}.
Furthermore let $M_{n}^{(s)}$ be the total number of observations produced by data source $s \! \in \! \Set{S}_{0}$ at time $n$.
These observations are represented by the vectors $\V{z}_{n,m}^{(s)} \in \mathbb{R}^{d_{\text{z}}^{(s)}}\!$, with $m \rmv \in \rmv \Set{M}_{n}^{(s)} \rmv \deq \{ 1, \ldots, M_{n}^{(s)} \}$.
We also define the
vectors $\V{z}_{n}^{(s)} \! \deq \! \big[\V{z}_{n,1}^{(s)\T}\rmv, \ldots, \linebreak \V{z}_{n,M_{n}^{(s)}}^{(s)\T}\big]^{\T}\rmv$,
$\V{z}_{n} \! \deq \rmv \big[\V{z}_{n}^{(0)\T}\rmv,\ldots,\V{z}_{n}^{(S)\T}\big]^{\T}\rmv$, and
$\V{z} \rmv \deq \rmv [\V{z}_{1}^{\T}\rmv,\ldots,\V{z}_{n}^{\T}]^{\T}\rmv$
as well as
$\V{m}_{n} \! \deq \rmv \big[ M_{n}^{(0)}\rmv,\ldots,M_{n}^{(S)} \big]^{\T}$ and 
$\V{m} \rmv\deq\rmv [\V{m}_{1}^{\T}\rmv, \ldots, \V{m}_{n}^{\T}]^{\T}\rmv$.

A TIPS sensor $s \! \in \! \Set{S}$ detects an existing PT $k$ at time $n$, in the sense that PT $k$ generates a measurement $\V{z}^{(s)}_{n,m}$, with probability $P_{\text{d}}^{(s)}(\V{x}_{n,k})$.
The dependence of $\V{z}_{n,m}^{(s)}$ on the PT state $\V{x}_{n,k}$  is modeled by the likelihood function $f \big( \V{z}_{n,m}^{(s)} \big| \V{x}_{n,k} \big)$.
Furthermore, we assume that the number of false alarms generated by TIPS sensor $s$ is Poisson distributed with mean $\mu^{(s)}$, and 
each false alarm is distributed according  to the pdf $f_{\text{FA}} \big(\V{z}_{n,m}^{(s)}\big)$.

For the TDRS, 
the observation vector $\V{z}_{n,m}^{(0)}$ represents the $m$th cluster at time $n$.
As mentioned in Section~\ref{sec:problem_formulation}, a
cluster consists of a group of reports with the same
RID $d \! \in \! \Set{D}$, or of an individual report with an
RID $d \! \notin \! \Set{D}$.
We denote by
$\zeta_{n,m} \in \Set{D} \cup \{ 0 \} = \{ 0,1,\ldots,D \}$
the RID of cluster $m$, 
where\footnote{Note that ``$\tau_{n,k} = 0$'' and ``$\zeta_{n,m} = 0$'' have different meanings: 
the former means that PT $k$ is noncooperative, whereas the latter means that the $m$th report received by the TDRS contains a corrupted
RID $d \! \notin \! \Set{D}$.}
$\zeta_{n,m} = 0$ identifies the case $d \! \notin \! \Set{D}$.
Moreover, we denote 
by $L_{n,m}$ the number of self-measurements within cluster $m$,
by $\V{q}_{n,m}^{(\ell)}$ with 
$\ell \! \in \! \Set{L}_{n,m} \! \deq\! \{1, \ldots, L_{n,m} \}$ 
the $\ell$th self-measurement within cluster $m$, and by $\V{q}_{n,m} \! \deq \! \big[\V{q}_{n,m}^{(1)\T},\ldots,\V{q}_{n,m}^{(L_{n,m})\T}\big]^{\T}$ 
the vector comprising all the self-measurements in cluster $m$.
The observation vector $\V{z}_{n,m}^{(0)}$ is thus given by
$\V{z}_{n,m}^{(0)} \deq \big[\V{q}_{n,m}^{\T},\zeta_{n,m}\big]^{\T}$.
Each self-measurement 
$\V{q}_{n,m}^{(\ell)}$ is generated by a
cooperative PT $k$ at some intermediate time $t_{n,m}^{(\ell)} \in ( t_{n-1},t_{n} ]$. 
(We use the convention that $\ell < \ell'$ implies $t_{n,m}^{(\ell)} \leqslant t_{n,m}^{(\ell')}$.)
The dependence of $\V{q}_{n,m}^{(\ell)}$ on 
the state of cooperative PT $k$
is modeled
by the likelihood function $f \big(\V{q}_{n,m}^{(\ell)} \big| \V{x}_{n,k,m}^{(\ell)} \big)$,
where $\V{x}_{n,k,m}^{(\ell)}$ is the state of cooperative PT $k$ at time $t_{n,m}^{(\ell)}$.

Next, we consider the likelihood function $f\big( \V{z}_{n,m}^{(0)} \big| \V{x}_{n,k}, \linebreak \tau_{n,k} \big)$, which describes the statistical dependency of
$\V{z}_{n,m}^{(0)} \rmv= \big[\V{q}_{n,m}^{\T},\zeta_{n,m}\big]^{\T}$ on
$\V{x}_{n,k}$
and
$\tau_{n,k}$.
We assume that (i) 
given the cooperative PT state $\V{x}_{n,k}$, $\V{q}_{n,m}$ is conditionally independent of the TID $\tau_{n,k}$ and the RID $\zeta_{n,m}$, and
(ii) given the TID $\tau_{n,k}$, the RID $\zeta_{n,m}$ is conditionally independent of $\V{x}_{n,k}$.
With these assumptions, we 
\vspace{-.2mm}
obtain
\begin{align}
	\nonumber \\[-6mm]
	f\big( \V{z}_{n,m}^{(0)} \big| &\V{x}_{n,k}, \tau_{n,k} \big) = f\big( \V{q}_{n,m}, \zeta_{n,m} \big| \V{x}_{n,k}, \tau_{n,k} \big) \nonumber \\[0mm]
	&=f(\V{q}_{n,m} | \zeta_{n,m}, \V{x}_{n,k}, \tau_{n,k} ) \ist p(\zeta_{n,m} | \V{x}_{n,k}, \tau_{n,k} ) \nonumber\\[0mm]
	&= f(\V{q}_{n,m} | \V{x}_{n,k}) \ist p(\zeta_{n,m} | \tau_{n,k}).
	\label{eq:factorisation_likelihood_ais_single_cluster} \\[-5.7mm]
	\nonumber
\end{align}
The likelihood function $f(\V{q}_{n,m} | \V{x}_{n,k})$, 
which was not reported in our previous works \cite{gaglione18,SoldiGMHBFW:C19}, is 
derived
in the supplementary material manuscript \cite{SupplMat}; this 
derivation
involves the previously introduced function $f \big(\V{q}_{n,m}^{(\ell)} \big| \V{x}_{n,k,m}^{(\ell)} \big)$. 
For an expression of the factor
$p(\zeta_{n,m} | \tau_{n,k})$,
we note
the following facts about the RID $\zeta_{n,m}$ of a report received by the TDRS and transmitted by a cooperative PT $k$ with TID
$\tau_{n,k}
\! \in \! \Set{D}$:
(i) $\zeta_{n,m}$ coincides with $\tau_{n,k}$ with probability $p_{n,m}^{\text{c}}$;
(ii) it is corrupted and does not belong to $\Set{D}$
with probability $p_{n,m}^{\text{e}}$; and
(iii) it is corrupted and belongs to $\Set{D}$ but is different from $\tau_{n,k}$
with probability $(1 - p_{n,m}^{\text{c}} - p_{n,m}^{\text{e}})/(D - 1)$.
That is,
\linebreak
\begin{align}
	\nonumber \\[-10mm]
	&p(\zeta_{n,m} | \tau_{n,k}) = 
	\begin{dcases}
		p_{n,m}^{\text{c}} \ist, & \!\!\zeta_{n,m} \!=\rmv \tau_{n,k}, \\[0mm]
		p_{n,m}^{\text{e}} \ist, & \!\!\zeta_{n,m} \!=\rmv 0, \\[0mm]
		\frac{ 1 \!-\rmv p_{n,m}^{\text{c}} \!-\rmv p_{n,m}^{\text{e}} }{D \rmv-\! 1} \ist, & \!\!\zeta_{n,m} \!\in\rmv \Set{D} \rmv\setminus\rmv \{ \tau_{n,k} \} \ist ,
	\end{dcases}
	\label{eq:likelihood_AIS_label}
\end{align}
for $\tau_{n,k} \neq 0$.
(For $\tau_{n,k} = 0$, i.e., for noncooperative PTs, the pmf $p(\zeta_{n,m} | \tau_{n,k})$ is not defined since noncooperative PTs do not send TDRS reports.)
We note that $p_{n,m}^{\text{c}}$ is the probability that the PT-to-TDRS transmission
is successful and without errors, 
and $p_{n,m}^{\text{e}}$ is the probability that there is an error
resulting in an RID $\zeta_{n,m} \!\notin\rmv \Set{D}$.
In particular, setting $p_{n,m}^{\text{c}} + p_{n,m}^{\text{e}} = 1$ excludes the possibility of a transmission error resulting in an RID $\zeta_{n,m} \in \Set{D}$ 
that differs from the transmitted TID $\tau_{n,k}$.

\vspace{-1mm}

\subsection{
Observation-Origin Uncertainty}
\label{subsec:target_observation_association}

The TIPS and TDRS observations have an uncertain origin, since it is not known from which PTs they originate, and TIPS observations may also be false alarms. 
To model this
observation-origin uncertainty, we make
two assumptions. For any data source $s \! \in \! \Set{S}_{0}$, the \textit{point-target assumption} 
\cite{Mah:B07,BarWilTia:B11} states that at each time step and at each data source, an existing PT can generate at most one observation and an observation can originate from
at most one existing PT. Furthermore, the \textit{no-false-alarms assumption} states that the TDRS ($s \rmv=\rmv 0$) cannot produce false alarms, i.e.,
each TDRS observation
must originate from an existing
cooperative PT.\footnote{From now on, we will 
speak of ``cooperative PTs'' instead of ``existing cooperative PTs.''
Indeed, we recall from Section~\ref{sec:state} that a nonexisting PT is noncooperative; as a consequence, a cooperative PT necessarily exists, i.e., $\tau_{n,k} 
\in \Set{D}$ implies $r_{n,k} = 1$.}
This implies that, at each time step $n$, the number of
cooperative PTs, denoted $N(\V{\tau}_{n})$,
cannot be smaller than the number of TDRS clusters, 
i.e., $M_{n}^{(0)} \! \leqslant \! N(\V{\tau}_{n}) \rmv$.
Note that $N(\V{\tau}_{n}) = \sum_{k = 1}^{K} (1 - \mathds{1}(\tau_{n,k}))$.
An association between the $K$ PTs and the $M_{n}^{(s)}$ observations produced by data source $s \! \in \! \Set{S}_{0}$ at time $n$ will be called \emph{admissible}
if it satisfies the above two assumptions.

The PT-observation association
for data source $s \! \in \! \Set{S}_{0}$ at time $n$ can be described by the 
\textit{PT-oriented association vector} $\V{a}_{n}^{(s)} \! \deq \rmv \big[a_{n,1}^{(s)},\ldots,a_{n,K}^{(s)}\big]^{\T}$ and alternatively by the 
\textit{observation-oriented association vector} $\V{b}_{n}^{(s)}  \! \deq \rmv \big[b_{n,1}^{(s)},\ldots,b_{n,M_{n}^{(s)}}^{(s)}\big]^{\T}$ 
\cite{WilLau:J14,MeyKroWilLauHlaBraWin:J18}.
Here, $a_{n,k}^{(s)}$ is defined as $m \! \in \! \Set{M}_{n}^{(s)}$ if PT $k$ generates observation $m$ and $0$ if PT $k$ does not generate any observation, and
$b_{n,m}^{(s)}$ is defined as $k \! \in \! \Set{K}$ if observation $m$ originates from PT $k$ and $0$ if observation $m$ does not originate from a PT.
We also define the vectors  $\V{a}_{n} \rmv \deq \rmv \big[\V{a}_{n}^{(0)\T}\rmv,\ldots,\V{a}_{n}^{(S)\T}\big]^{\T}$,
$\V{b}_{n} \rmv \deq \rmv \big[\V{b}_{n}^{(0)\T}, \linebreak \ldots,\V{b}_{n}^{(S)\T}\big]^{\T}\rmv$,
$\V{a} \rmv \deq \rmv \big[\V{a}_{1}^{\T},\ldots,\V{a}_{n}^{\T} \big]^{\T}$, and $\V{b} \rmv \deq \rmv \big[ \V{b}_{1}^{\T},\ldots,\V{b}_{n}^{\T} \big]^{\T}\rmv$.

The alternative descriptions of a PT-observation association 
that are provided by the association vectors $\V{a}_{n}^{(s)}$ and $\V{b}_{n}^{(s)}$ 
are equivalent, because, after $M_{n}^{(s)}$ is observed, $\V{a}_{n}^{(s)}$ can be derived directly from $\V{b}_{n}^{(s)}$ and vice versa.
However, using \emph{both} $\V{a}_{n}^{(s)}$ and $\V{b}_{n}^{(s)}$ in parallel makes it possible to mathematically characterize the admissibility of any PT-observation association
via an indicator function $\Psi^{(s)}\big( \V{a}_{n}^{(s)} \rmv, \V{b}_{n}^{(s)} \big)$ that factors into $K M_{n}^{(s)}$ component indicator functions. 
More specifically,
we define $\Psi^{(s)}\big( \V{a}_{n}^{(s)}\rmv,\V{b}_{n}^{(s)} \big)$ to be $1$ if $\V{a}_{n}^{(s)}$ and $\V{b}_{n}^{(s)}$ 
describe the same admissible association event, and to be $0$ otherwise. Then 
\begin{align}
	\nonumber \\[-6.5mm]
	\Psi^{(s)}\big( \V{a}_{n}^{(s)}\rmv,\V{b}_{n}^{(s)} \big) = \prod_{k = 1}^{K}\prod_{m = 1}^{M_{n}^{(s)}} \!\psi^{(s)}\big( a_{n,k}^{(s)},b_{n,m}^{(s)} \big),
	\label{eq:psi_ais_two_arguments}
\end{align}
where the factors $\psi^{(s)}\big( a_{n,k}^{(s)},b_{n,m}^{(s)} \big)$ are given as follows. For $s \! \in \! \Set{S}$
(TIPS sensor),
$\psi^{(s)}\big( a_{n,k}^{(s)}, b_{n,m}^{(s)} \big)$ expresses the point-target assumption and is thus $0$ if $a_{n,k}^{(s)} \!=\! m$ and $b_{n,m}^{(s)} \!\neq\! k$ or if \linebreak $a_{n,k}^{(s)} \!\neq\! m$ and $b_{n,m}^{(s)} \!=\! k$, and $1$ otherwise.
For $s \! = \! 0$
(TDRS), $\psi^{(0)}\big( a_{n,k}^{(0)},b_{n,m}^{(0)} \big)$
also
takes into account the no-false-alarms assumption, and thus
$\psi^{(0)}\big( a_{n,k}^{(0)},b_{n,m}^{(0)} \big)$ is $0$ if $a_{n,k}^{(0)} \!=\! m$ and $b_{n,m}^{(0)} \!\neq\! k$ or if $a_{n,k}^{(0)} \!\neq\! m$ and $b_{n,m}^{(0)} \!=\! k$ or if $b_{n,m}^{(0)} \!=\! 0$, and $1$ otherwise.
Note that each component indicator function $\psi^{(s)}\big( a_{n,k}^{(s)},b_{n,m}^{(s)} \big)$
corresponds to one individual PT-observation association, i.e., to one pair $(k,m)$ with $k \! \in \! \Set{K}$ and $m \! \in \! \Set{M}_{n}^{(s)}\rmv$.
The factorization \eqref{eq:psi_ais_two_arguments} is
key as it enables the development of scalable message passing algorithms \cite{WilLau:J14,MeyBraWilHla:J17,MeyKroWilLauHlaBraWin:J18}.

\vspace{-1mm}

\subsection{Joint Prior Distribution} \label{sec:joint}

Next, we derive a factorization of the joint prior distribution of the association variables $\V{a}$ and $\V{b}$, the numbers of observations $\V{m}$, 
and the augmented PT states $\V{y}$, i.e., $f (\V{a},\V{b},\V{m},\V{y})$. Because
$f(\V{a},\V{b},\V{m},\V{y}) \! = \! p(\V{a},\V{b},\V{m}|\V{y}) \ist f(\V{y})$,
with $f (\V{y})$ given by \eqref{eq:joint_pdf_augmented_state},
it remains to consider $p (\V{a},\V{b},\V{m} | \V{y})$. 
Assuming that $\V{a}_{n}^{(s)}\rmv $, $\V{b}_{n}^{(s)}\rmv $, and $M_{n}^{(s)}$
are conditionally independent,
given $\V{y}$, across time $n$ and data source index $s \! \in \! \Set{S}_{0}$ \cite{Mah:B07,BarWilTia:B11}, we obtain 
\begin{align}
	\hspace*{-5mm}p (\V{a},\V{b},\V{m} | \V{y})
	& =\prod_{n' = 1}^{n} \! p\big( \V{a}_{n'}^{(0)}\rmv ,\V{b}_{n'}^{(0)}\rmv , M_{n'}^{(0)} \big| \V{y}_{n'} \big) \nonumber\\[-0mm]
	& \hspace{4mm} \times \prod_{s = 1}^{S} p\big( \V{a}_{n'}^{(s)}\rmv ,\V{b}_{n'}^{(s)}\rmv ,M_{n'}^{(s)} \big| \V{y}_{n'} \big) \ist.
	\label{eq:prior_da_factorisation} \\[-6mm]
	\nonumber
\end{align}

For the TDRS, i.e., $s \! = \! 0$,
the pmf $p\big(\V{a}_{n}^{(0)}\rmv, \V{b}_{n}^{(0)}\rmv, M_{n}^{(0)} \big| \V{y}_{n}\big)$ ca be factorized by using the chain rule as follows:
\begin{align}
	& \hspace{0mm} p\big(\V{a}_{n}^{(0)}\rmv, \V{b}_{n}^{(0)}\rmv, M_{n}^{(0)} \big| \V{y}_{n}\big) 
	= p\big(\V{b}_{n}^{(0)} \big| \V{a}_{n}^{(0)}\rmv, M_{n}^{(0)} \rmv, \V{y}_{n}\big) \nonumber \\[.8mm]
	& \hspace{10mm} \times p\big(\V{a}_{n}^{(0)} \big| M_{n}^{(0)} \rmv, \V{y}_{n}\big) \ist p\big( M_{n}^{(0)} \big| \V{y}_{n}\big) \ist.
	\label{eq:prior_da_ais_factorisation}
\end{align}
Let us consider the three factors in this expression. For the first factor, since $\V{b}_{n}^{(0)}$ is fully described by $\V{a}_{n}^{(0)}\rmv$ and $M_{n}^{(s)}\rmv$, we have
$p\big(\V{b}_{n}^{(0)} \big| \V{a}_{n}^{(0)}\rmv, M_{n}^{(0)} \rmv, \V{y}_{n}\big) = p\big(\V{b}_{n}^{(0)} \big| \V{a}_{n}^{(0)}\rmv, M_{n}^{(s)} \big)$.
Then, using the indicator function $\Psi^{(0)}\big( \V{a}_{n}^{(0)}\rmv,\V{b}_{n}^{(0)} \big)$, which is
$1$ if $\V{a}_{n}^{(s)}$ and $\V{b}_{n}^{(s)}$ 
describe the same admissible association event
and $0$ otherwise, we can write $p\big(\V{b}_{n}^{(0)} \big| \V{a}_{n}^{(0)}\rmv, M_{n}^{(s)} \big) = \Psi^{(0)}\big( \V{a}_{n}^{(0)}\rmv,\V{b}_{n}^{(0)} \big)$.
To obtain the second factor,
$p\big(\V{a}_{n}^{(0)} \big| M_{n}^{(0)} \rmv, \V{y}_{n}\big)$, we observe that, given $M^{(0)}_{n}\rmv$ and $\V{y}_{n}$, the number of PT-cluster associations that satisfy the point-target and no-false-alarms assumptions is equivalent to the number of draws of $M^{(0)}_{n}$ PTs out of the $N(\V{\tau}_{n})$ 
cooperative PTs,
where the draws are without replacement and with the drawing order respected.
Assuming each of these draws equally likely, it can be shown that
\begin{align}
	&\hspace*{0mm} p\big(\V{a}_{n}^{(0)} \big| M_{n}^{(0)} \rmv, \V{y}_{n}\big) = p\big(\V{a}_{n}^{(0)} \big| M_{n}^{(0)} \rmv, \V{x}_{n}, \V{r}_{n}, \V{\tau}_{n} \big) \nonumber\\[.5mm]
	&\hspace{4mm} = \frac{\big(N(\V{\tau}_{n}) \rmv-\! M_{n}^{(0)}\big)!}{N(\V{\tau}_{n})!} \prod_{k = 1}^{K}
	\lambda (r_{n,k}, \tau_{n,k}, a_{n,k}^{(0)}) \ist .
	\label{eq:prior_da_ais_a} \\[-6mm]
	\nonumber
\end{align}
Here, the function $\lambda(r_{n,k},\tau_{n,k},a_{n,k}^{(0)})$ is defined as $1$ if $r_{n,k} = 1$ and $\tau_{n,k} \neq 0$, and as $\mathds{1}(a_{n,k}^{(0)})$ otherwise.
This function ensures that a PT that is nonexisting
($r_{n,k} = 0$) or existing and noncooperative
($r_{n,k} = 1$ and $\tau_{n,k} = 0$) cannot be associated to a TDRS cluster.
Finally, the third factor in \eqref{eq:prior_da_ais_factorisation}, $p\big( M_{n}^{(0)} \big| \V{y}_{n} \big) = p\big( M_{n}^{(0)} \big| \V{x}_{n}, \V{r}_{n}, \V{\tau}_{n} \big)$,
expresses the prior distribution of the number of TDRS clusters.
In the absence of
further knowledge, we assume that $M_{n}^{(0)}$ is uniform between $0$ and $N(\V{\tau}_{n})$, i.e.,
$p\big( M_{n}^{(0)} \big| \V{x}_{n}, \V{r}_{n}, \V{\tau}_{n} \big) = 1/\big( N(\V{\tau}_{n}) + 1 \big)$ if $M_{n}^{(0)} \leqslant N(\V{\tau}_{n})$
and $0$ otherwise.
Thus, by inserting
\eqref{eq:prior_da_ais_a} into \eqref{eq:prior_da_ais_factorisation}, and using \eqref{eq:psi_ais_two_arguments}, we obtain
\begin{align}
	&p\big(\V{a}_{n}^{(0)}\rmv, \V{b}_{n}^{(0)}\rmv, \ist M_{n}^{(0)} \big| \V{y}_{n} \big) = p\big(\V{a}_{n}^{(0)}\rmv, \V{b}_{n}^{(0)}\rmv, \ist M_{n}^{(0)} \big| \V{x}_{n}, \V{r}_{n}, \V{\tau}_{n} \big) \nonumber \\[.8mm]
	&\hspace{1mm} = \chi\big( \V{\tau}_{n} ,M_{n}^{(0)} \big) \prod_{k = 1}^{K} \rmv
	\lambda (r_{n,k}, \tau_{n,k}, a_{n,k}^{(0)}) \prod_{m = 1}^{M_{n}^{(0)}} \!\psi^{(0)}\big( a_{n,k}^{(0)},b_{n,m}^{(0)} \big) ,
	\nonumber \\[-2.5mm]
	\label{eq:prior_da_ais_factorisation_final}\\[-10mm]
\nonumber
\end{align}
where
\vspace{-.5mm}
\begin{align}
	&\hspace{-7mm}\chi\big( \V{\tau}_{n} ,M_{n}^{(0)} \big) \deq
	\begin{dcases}
		\frac{\big( N(\V{\tau}_{n}) \rmv-\! M_{n}^{(0)}\big)!}{ \big( N(\V{\tau}_{n}) + 1 \big) !} \ist,
		& \!\!M_{n}^{(0)} \!\leqslant\! N(\V{\tau}_{n})	\\[-1mm]
		0 \ist, & \!\!\text{otherwise.}
	\end{dcases}
	\label{eq:chi_factor} \\[-6mm]
	\nonumber
\end{align}
Note that expression \eqref{eq:prior_da_ais_factorisation_final} does not depend on $\V{x}_{n,k}$.

For a TIPS sensor $s \! \in \! \Set{S}$,
the joint prior distribution $p\big(\V{a}_{n}^{(s)}\rmv, \linebreak \V{b}_{n}^{(s)}\rmv, M_{n}^{(s)} \big| \V{y}_{n} \big)$ was
derived in \cite{MeyBraWilHla:J17}, 
and is reported here for completeness:
\vspace{-1mm}
\begin{align}
	p\big(\V{a}_{n}^{(s)}\rmv, \V{b}_{n}^{(s)}\rmv, \ist M_{n}^{(s)} \big| \V{y}_{n} \big) &= C\big(M_{n}^{(s)}\big) 
	   \prod_{k = 1}^{K}  \rmv h^{(s)} \big( \V{y}_{n,k},a_{n,k}^{(s)}; M_{n}^{(s)} \big)\nonumber\\[-1.2mm]
	& \hspace{4mm} \times\rmv \prod_{m = 1}^{M_{n}^{(s)}} \!\psi^{(s)}\big( a_{n,k}^{(s)},b_{n,m}^{(s)} \big) \ist .
	\label{eq:prior_da_radar_factorisation_final}  \\[-5.5mm]
	\nonumber
\end{align}
Here, $C\big(M_{n}^{(s)}\big) \! \deq \! e^{-\mu^{(s)}} (\mu^{(s)})^{M_{n}^{(s)}} / M_{n}^{(s)}!$, 
and $h^{(s)} \big( \V{y}_{n,k},\linebreak a_{n,k}^{(s)}; M_{n}^{(s)} \big) = h^{(s)} \big( \V{x}_{n,k}, r_{n,k}, \tau_{n,k}, a_{n,k}^{(s)}; M_{n}^{(s)} \big)$,
$s \! \in \! \Set{S}$, is defined for $r_{n,k} \! = \! 1$ as 
$P_{\text{d}}^{(s)}(\V{x}_{n,k}) / \mu^{(s)}$ if $a_{n,k}^{(s)} \! \in \! \Set{M}_{n}^{(s)}$ and 
$1 - P_{\text{d}}^{(s)}(\V{x}_{n,k})$ if $a_{n,k}^{(s)} \! = \! 0$, and for $r_{n,k} \! = \! 0$ as 
$\mathds{1} \big( a_{n,k}^{(s)} \big)$.
Note that 
$h^{(s)} \big( \V{y}_{n,k},a_{n,k}^{(s)}; M_{n}^{(s)} \big)$, $s \! \in \! \Set{S}$, 
does not depend on 
$\tau_{n,k}$.

Finally, by inserting \eqref{eq:prior_da_ais_factorisation_final} and \eqref{eq:prior_da_radar_factorisation_final} into 
\eqref{eq:prior_da_factorisation}, the overall pmf $p (\V{a},\V{b}, \V{m} | \V{y}) = p (\V{a},\V{b}, \V{m} | \V{x}, \V{r}, \V{\tau})$ becomes
\begin{align}
	p (\V{a},\V{b},\V{m} | \V{x}, \V{r}, \V{\tau}) &=\ist C_{\Set{M}} (\V{m}) \! \prod_{n' = 1}^{n} \! \chi\big( \V{\tau}_{n'} ,M_{n'}^{(0)} \big) \nonumber \\[-1mm]
	&\hspace{-7mm}\times \prod_{s = 0}^{S}\prod_{k = 1}^{K}  \rmv h^{(s)} \big( \V{x}_{n'\!,k},r_{n'\!,k},\tau_{n'\!,k},a_{n'\!,k}^{(s)}; M_{n'}^{(s)} \big) \nonumber \\[0mm]
	&\hspace{-7mm}\times\! \prod_{m = 1}^{M_{n'}^{(s)}} \!\psi^{(s)}\big( a_{n'\!,k}^{(s)},b_{n'\!,m}^{(s)} \big) \ist ,
	\label{eq:factorisation_association_variables_final} \\[-5mm]
	\nonumber
\end{align}
where $C_{\Set{M}}(\V{m}) \deq \prod_{n' = 1}^{n} \prod_{s = 1}^{S} C\big(M_{n'}^{(s)}\big)$ and
$h^{(0)} ( \V{x}_{n,k}, r_{n,k}, \linebreak \tau_{n,k}, a_{n,k}^{(0)}; M_{n}^{(0)})$ is defined for $r_{n,k} = 1$ as $1$ if $\tau_{n,k} \neq 0$ and $\mathds{1}(a_{n,k}^{(0)})$ if $\tau_{n,k} = 0$, and for $r_{n,k} = 0$ as $\mathds{1}(a_{n,k}^{(0)})$.
We observe that expression \eqref{eq:factorisation_association_variables_final} is defined only if $r_{n,k} = 0$ implies $\tau_{n,k} = 0$.
Indeed, for other choices of $\V{r}$ and $\V{\tau}$, the value of $p (\V{a},\V{b},\V{m} | \V{x}, \V{r}, \V{\tau})$ is irrelevant to the joint posterior distribution provided later in Section~\ref{sec:proposed_algorithm}, since these choices yield $p(\tau_{n,k} | r_{n,k}, r_{n-1,k}, \tau_{n-1,k}) = 0$ (cf. \eqref{eq:single_target_label_pdf_1}).
We emphasize that the formulation of the joint prior pmf in \eqref{eq:factorisation_association_variables_final} in terms of \emph{both} the association vectors
$\V{a}$ and $\V{b}$ allows the joint prior pmf to factorize with respect to both $k$ and $m$; this factorization, in turn, is key to developing a probabilistic data association scheme with 
reduced complexity \cite{MeyBraWilHla:J17,MeyKroWilLauHlaBraWin:J18,WilLau:J14}.

\vspace{-1mm}

\subsection{Likelihood Function}
\label{sec:global-likelihood}

The likelihood function $f (\V{z} | \V{y}, \V{a}, \V{b}, \V{m})$ expresses the statistical dependence of the observation $\V{z}$ on the 
augmented states $\V{y}$, the association vectors $\V{a}$ and $\V{b}$, and the number-of-observations vector $\V{m}$.
Since the information carried by $\V{a}$ and by $\V{b}$ is equivalent 
once $\V{m}$ is observed, we have $f (\V{z} | \V{y}, \V{a}, \V{b},\linebreak \V{m}) 
= f (\V{z} | \V{y}, \V{a}, \V{m})$.
Then, under the assumption that given $\V{y}$, $\V{a}$, and $\V{m}$, the observations $\V{z}_{n}^{(s)} \! \in \rmv \mathbb{R}^{d_{\text{z}}^{(s)} M_{n}^{(s)}}$ are conditionally independent across time $n$ and data source index 
$s \! \in \! \Set{S}_{0}$, the
likelihood function can be factorized as \cite{Mah:B07,BarWilTia:B11}
\begin{align}
	\nonumber \\[-6mm]
	 f(\V{z} | \V{y},\V{a},\V{m})
	& = \prod_{n' = 1}^{n} \! f\big(\V{z}_{n'}^{(0)} \big| \V{y}_{n'}, \V{a}_{n'}^{(0)}\rmv, M_{n'}^{(0)}\big) \nonumber \\[-1mm]
	&\hspace{7mm} \times \prod_{s = 1}^{S} f\big(\V{z}_{n'}^{(s)} \big| \V{y}_{n'}, \V{a}_{n'}^{(s)}\rmv, M_{n'}^{(s)}\big) \ist.
	\label{eq:factorisation_likelihood} \\[-6mm]
	\nonumber
\end{align}
Assuming moreover that the clusters $\V{z}_{n,m}^{(0)}$ are conditionally independent given
$\V{y}_{n}$, $\V{a}_{n}^{(0)}\rmv$, and $M_{n}^{(0)}\rmv$,
the TDRS likelihood function $f\big(\V{z}_{n}^{(0)} \big| \V{y}_{n}, \V{a}_{n}^{(0)}\rmv, M_{n}^{(0)}\big)$ can be written as
\begin{align}
	\hspace{-5mm} f\big(\V{z}_{n}^{(0)} \big| \V{y}_{n}, \V{a}_{n}^{(0)}\rmv, M_{n}^{(0)}\big) &= \prod_{m = 1}^{M_{n}^{(0)}} \! f\big( \V{z}_{n,m}^{(0)} \big| \V{y}_{n}, \V{a}_{n}^{(0)} \big) \nonumber\\
	&= \rmv\prod_{k \in \Set{K}_{n,\tiny{\V{a}}}^{(0)}} \!\!\rmv f \Big( \V{z}_{n,a_{n,k}^{(0)}}^{(0)} \Big| \V{x}_{n,k}, \tau_{n,k} \Big) 
	\label{eq:factorisation_likelihood_ais} \\[-6mm]
	\nonumber
\end{align}
if the dimension of the vector $\V{z}_{n}^{(0)}$ is consistent with $M_{n}^{(0)}$ 
in the sense that $\V{z}_{n}^{(0)} \! \in \mathbb{R}^{d_{\text{z}}^{(0)} M_{n}^{(0)}}\!$, and $f\big(\V{z}_{n}^{(0)} \big| \V{y}_{n}, \V{a}_{n}^{(0)}\rmv, M_{n}^{(0)}\big) = 0$ otherwise.
Here, $\Set{K}_{n,\scriptsize{\V{a}}}^{(0)} \rmv\deq\rmv \big\{ k \! \in \! \Set{K} \!: r_{n,k} \!=\! 1, \tau_{n,k} \!\neq\rmv 0, a_{n,k}^{(0)} \!\rmv = \! m \! \in \! \Set{M}_{n}^{(0)} \big\}$
is the set of
cooperative PTs 
that generate
a cluster of TDRS reports at time $n$.
We note that expression \eqref{eq:factorisation_likelihood_ais}
is defined
only if $r_{n,k} = 0$ implies $\tau_{n.k} = 0$, $\V{a}_{n}^{(0)}$ satisfies the point-target and no-false-alarms assumptions, and $M_{n}^{(0)} \leqslant N(\V{\tau}_{n})$.
Indeed, for other choices of $\V{y}_{n}$, $\V{a}_{n}^{(0)}\rmv$, or $M_{n}^{(0)}\rmv$, the value of $f\big(\V{z}_{n}^{(0)} \big| \V{y}_{n}, \V{a}_{n}^{(0)}\rmv, M_{n}^{(0)}\big)$ 
is irrelevant to the joint posterior distribution since these choices yield $p (\tau_{n,k} | r_{n,k},r_{n-1,k}, \tau_{n-1,k}) = 0$ (cf. \eqref{eq:single_target_label_pdf_1}) 
or $p\big(\V{a}_{n}^{(0)}\rmv, \V{b}_{n}^{(0)}\rmv, M_{n}^{(0)} \big| \V{x}_{n}, \V{r}_{n}, \V{\tau}_{n}\big) = 0$ (cf. \eqref{eq:prior_da_ais_factorisation_final}).
By extending the product in \eqref{eq:factorisation_likelihood_ais} to all $k \! \in \! \Set{K}$, the TDRS likelihood function can be expressed 
as
\begin{align}
	\nonumber \\[-6mm]
	\hspace{-3.5mm} f\big(\V{z}_{n}^{(0)} \big| \V{y}_{n}, \V{a}_{n}^{(0)}\rmv, M_{n}^{(0)}\big) = \prod_{k = 1}^{K} g^{(0)} \big(\V{y}_{n,k},a_{n,k}^{(0)}; \V{z}_{n}^{(0)} \big)\ist,
	\label{eq:factorisation_likelihood_ais_with_function_g} \\[-6mm]
	\nonumber
\end{align}
with $g^{(0)} \big(\V{y}_{n,k},a_{n,k}^{(0)}; \V{z}_{n}^{(0)} \big) \!=\! g^{(0)} \big(\V{x}_{n,k}, r_{n,k}, \tau_{n,k}, a_{n,k}^{(0)};\V{z}_{n}^{(0)} \big)$
defined for $r_{n,k} \! = \! 1$ and $\tau_{n,k} \neq 0$ as 
$f\big(\V{z}_{n,m}^{(0)} \big| \V{x}_{n,k},\tau_{n,k} \big)$ (cf. \eqref{eq:factorisation_likelihood_ais_single_cluster}) if $a_{n,k}^{(0)} \! = \! m \! \in \! \Set{M}_{n}^{(0)}$ and 
$1$ if $a_{n,k}^{(0)} \! = \! 0$, and for $r_{n,k} \! = \! 0$ or $\tau_{n,k} = 0$ as 
$1$.

For the TIPS
sensors, the likelihood function $f\big(\V{z}_{n}^{(s)} \big| \V{y}_{n}, \linebreak \V{a}_{n}^{(s)}\rmv, M_{n}^{(s)}\big)$, $s \! \in \! \Set{S}$, is given by  \cite{MeyBraWilHla:J17}
\begin{align}
	\nonumber \\[-6mm]
	f\big(\V{z}_{n}^{(s)} \big| \V{y}_{n}, \V{a}_{n}^{(s)}\rmv, M_{n}^{(s)}\big) = C\big( \V{z}_{n}^{(s)} \big) \prod_{k = 1}^{K} g^{(s)} \big(\V{y}_{n,k},a_{n,k}^{(s)}; \V{z}_{n}^{(s)} \big) \nonumber \\[-3mm]
	\label{eq:factorisation_likelihood_radar_with_function_g} \end{align}
if the dimension of the vector $\V{z}_{n}^{(s)}$ is consistent with $M_{n}^{(s)}$, i.e., $\V{z}_{n}^{(s)} \!\in\rmv \mathbb{R}^{d_{\text{z}}^{(s)} M_{n}^{(s)}}\!$, 
and $f\big(\V{z}_{n}^{(s)} \big| \V{y}_{n}, \V{a}_{n}^{(s)}\rmv, M_{n}^{(s)}\big) = 0$ otherwise. Here,
$C\big( \V{z}_{n}^{(s)} \big) \rmv \deq \prod_{m = 1}^{M_{n}^{(s)}} f_{\text{FA}}(\V{z}_{n,m}^{(s)})$, 
and 
$g^{(s)} \big(\V{y}_{n,k},a_{n,k}^{(s)}; \V{z}_{n}^{(s)} \big)$\linebreak $= \! g^{(s)} \big(\V{x}_{n,k}, r_{n,k}, \tau_{n,k},a_{n,k}^{(s)}; \V{z}_{n}^{(s)} \big)$, $s \! \in \! \Set{S}$, 
is defined for $r_{n,k} \! = \! 1$ as $f \big( \V{z}_{n,m}^{(s)} \big| \V{x}_{n,k} \big) / f_{\text{FA}} \big( \V{z}_{n,m}^{(s)} \big)$ if $a_{n,k}^{(s)} \! = \! m \! \in \! \Set{M}_{n}^{(s)}$ and 
$1$ if $a_{n,k}^{(s)} \! = \! 0$, and for $r_{n,k} \! = \! 0$ as $1$. Note that 
$g^{(s)} \big(\V{y}_{n,k}, a_{n,k}^{(s)}; \V{z}_{n}^{(s)} \big)$, $s \! \in \! \Set{S}$, does not depend on $\tau_{n,k}$.
Similarly to \eqref{eq:factorisation_likelihood_ais}, expression \eqref{eq:factorisation_likelihood_radar_with_function_g} is defined only if $r_{n,k} = 0$ implies $\tau_{n.k} = 0$
and $\V{a}_{n}^{(s)}$ satisfies the point-target assumption. Indeed, for other choices of $\V{y}_{n}$ or $\V{a}_{n}^{(s)}\rmv$, the value of 
$f\big(\V{z}_{n}^{(s)} \big| \V{y}_{n}, \V{a}_{n}^{(s)}\rmv, M_{n}^{(s)}\big)$, $s \! \in \! \Set{S}$ is irrelevant to the joint posterior distribution 
since these choices yield $p (\tau_{n,k} | r_{n,k},r_{n-1,k}, \tau_{n-1,k}) = 0$ (cf. \eqref{eq:single_target_label_pdf_1}) or
$p\big(\V{a}_{n}^{(s)}\rmv, \V{b}_{n}^{(s)}\rmv, M_{n}^{(s)} \big| \linebreak \V{x}_{n}, \V{r}_{n}, \V{\tau}_{n}\big) = 0$ (cf. \eqref{eq:prior_da_radar_factorisation_final}).

Finally, by inserting \eqref{eq:factorisation_likelihood_ais_with_function_g} and \eqref{eq:factorisation_likelihood_radar_with_function_g} into \eqref{eq:factorisation_likelihood}, 
we obtain the
\vspace{-1.5mm}
likelihood 
function 
\begin{align}
	&f(\V{z} | \V{y},\V{a},\V{m}) = C_{\Set{S}}(\V{z}) \rmv\prod_{n' = 1}^{n} \prod_{s = 0}^{S} \prod_{k = 1}^{K} g^{(s)}\big( \V{y}_{n'\!,k},a_{n'\!,k}^{(s)}; \V{z}_{n'}^{(s)} \big) \ist, \nonumber\\[-3mm]
	\label{eq:factorisation_likelihood_with_function_g} \\[-7.5mm]
\nonumber
\end{align}
with $C_{\Set{S}}(\V{z}) \deq \prod_{n' = 1}^{n} \prod_{s = 1}^{S} C\big(\V{z}_{n'}^{(s)}\big)$.

\section{Posterior Distribution and Factor Graph}
\label{sec:proposed_algorithm}

We next consider
the joint posterior pdf $f (\V{y},\V{a},\V{b} | \V{z})$, which is needed for PT 
existence confirmation and PT state estimation. Indeed,
the ultimate objective of the proposed algorithm is to determine the existence of the PTs $k \!\in\! \Set{K}$ and to estimate the PTs' states $\V{x}_{n,k}$ and
TIDs $\tau_{n,k}$ from all the current and past observations, i.e., from $\V{z}$.
We will confirm that
PT $k$
exists
if its posterior existence probability $p(r_{n,k} \!=\! 1 | \V{z})$ is above a threshold $P_{\text{th}}$ \cite[Ch. 2]{Poo:B94}.
If confirmed, then estimates of its state and
TID are obtained 
as $\hat{\V{x}}_{n,k} \! = \rmv \int \rmv \V{x}_{n,k} \ist f(\V{x}_{n,k} | r_{n,k} \!=\! 1, \V{z}) \ist d\V{x}_{n,k}$
and $\hat{\tau}_{n,k} \! = \! \argmax_{\,\tau_{n,k} \in \Set{D}_{0}} \ist p(\tau_{n,k} | r_{n,k} \!=\! 1, \V{z}) \ist$, respectively \cite[Ch. 4]{Poo:B94}, 
where $\Set{D}_{0} \deq \Set{D} \cup \{ 0 \} = \{0,1, \ldots, D \}$. 
Here, the pmfs $p (r_{n,k} | \V{z})$ and $p (\tau_{n,k} | r_{n,k} \! = \! 1, \V{z})$ and the pdf $f (\V{x}_{n,k} | r_{n,k} \! = \! 1, \V{z})$ can be calculated by
simple elementary operations---including marginaliza\-tions---from $f (\V{x}_{n,k}, r_{n,k}, \linebreak \tau_{n,k} | \V{z})$, which, in turn, is a marginal density
of the joint posterior pdf $f (\V{y},\V{a},\V{b} | \V{z})$.
To obtain an expression of $f (\V{y},\V{a},\V{b} | \V{z})$, we recall that, once $\V{m}$ is observed,
the information carried by $\V{a}$ and by $\V{b}$ is equivalent, and we use Bayes' rule.
That is,
\begin{align}
	f (\V{y},\V{a},\V{b} | \V{z} ) &= \sum_{\V{m'}} f(\V{y},\V{a},\V{b},\V{m}' | \V{z}) \nonumber \\[.5mm]
	&\propto \sum_{\V{m'}} f(\V{z} | \V{y},\V{a},\V{b},\V{m}') \ist p(\V{a},\V{b},\V{m}' | \V{y}) \ist f(\V{y}) \nonumber \\[.5mm]
	&= \sum_{\V{m'}} f(\V{z} | \V{y},\V{a},\V{m}') \ist p(\V{a},\V{b},\V{m}' | \V{y}) \ist f(\V{y}) \ist .
	\label{eq:pdf_factorisation_old}
	\\[-6.5mm] \nonumber
\end{align}
Here, $ \sum_{\V{m'}}$ is the summation over all elements in 
$\mathbb{N}^{(S + 1)n}_0\rmv$.
Recalling the expression of $f(\V{z} | \V{y},\V{a},\V{m})$ in \eqref{eq:factorisation_likelihood} and the fact that 
$f\big(\V{z}_{n}^{(s)} \big| \V{y}_{n}, \V{a}_{n}^{(s)}\rmv, M_{n}^{(s)}\big)$, $s \! \in \! \Set{S}_{0}$ 
involved in \eqref{eq:factorisation_likelihood} 
is nonzero only if $\V{z}_{n}^{(s)}$ is consistent with $M_{n}^{(s)}\rmv$, we obtain further
\begin{align}
	f (\V{y},\V{a},\V{b} | \V{z} ) \propto 
f(\V{z} | \V{y},\V{a},\V{m}) \ist p(\V{a},\V{b},\V{m} | \V{y}) \ist f(\V{y}) \ist.
	\label{eq:pdf_factorisation}
\end{align}
Then, by inserting the expressions \eqref{eq:factorisation_likelihood_with_function_g} for $f (\V{z} | \V{y},\V{a},\V{m})$, \eqref{eq:factorisation_association_variables_final} 
for $p (\V{a},\V{b},\V{m} | \V{y})$, and \eqref{eq:joint_pdf_augmented_state}
for $f (\V{y})$, 
we finally obtain
\vspace{-1mm}
\begin{align}
	f (\V{y},\V{a},\V{b} | \V{z} ) &\propto \Bigg( \prod_{k' = 1}^{K} \! f(\V{y}_{0,k'}) \rmv\Bigg) \rmv \prod_{n' = 1}^{n} \!\chi\big(\V{\tau}_{n'},M_{n'}^{(0)}\big) \nonumber \\[0mm]
	&\hspace*{-5mm} \times\prod_{k = 1}^{K} \rmv f(\V{y}_{n'\!,k} | \V{y}_{n' \rmv- 1,k}) \prod_{s = 0}^{S} \upsilon^{(s)}\big(\V{y}_{n'\!,k},a_{n'\!,k}^{(s)}; \V{z}_{n'}^{(s)}\big) \nonumber \\[.5mm]
	& \hspace{-5mm} \times\! \prod_{m = 1}^{M_{n'}^{(s)}} \!\psi^{(s)}\big( a_{n'\!,k}^{(s)},b_{n'\!,m}^{(s)} \big) \ist ,
	\label{eq:pdf_factorisation_final} \\[-7mm]
\nonumber
\end{align}
where
\begin{align}
	\upsilon^{(s)} \rmv\big( \V{y}_{n,k},a_{n,k}^{(s)}; \V{z}_{n}^{(s)} \big) &\deq h^{(s)} \rmv \big(\V{y}_{n,k},a_{n,k}^{(s)}; M_{n}^{(s)}\big) \\
	&\hspace{4mm} \times \rmv g^{(s)} \rmv \big(\V{y}_{n,k},a_{n,k}^{(s)}; \V{z}_{n}^{(s)} \big) \ist.
	\label{eq:function_v}
\end{align}
Note that since $h^{(s)}(\V{y}_{n,k}, a_{n,k}^{(s)}; M_{n}^{(s)})$,
$s\! \in \! \Set{S}$ and $g^{(s)}(\V{y}_{n,k}, \linebreak a_{n,k}^{(s)}; \V{z}_{n}^{(s)})$, $s\! \in \! \Set{S}$
do not depend on $\tau_{n,k}$, it follows that $\upsilon^{(s)} \big( \V{y}_{n,k}, a_{n,k}^{(s)}; \V{z}_{n}^{(s)} \big)$, $s \! \in \! \Set{S}$ does not depend on $\tau_{n,k}$.

The factorization \eqref{eq:pdf_factorisation_final} is represented by the factor graph shown for one time step $n$ in Fig.~\ref{fig:factor_graph}.
This factor graph contains three different types of loops for each time step $n$.
Within each block corresponding to a data source $s \! \in \! \Set{S}_{0}$, there are ``inner'' loops 
across association variables $a_{n,k}^{(s)}$, $k \! \in \! \Set{K}$ and $b_{n,m}^{(s)}$, $m \! \in \! \Set{M}_{n}^{(s)}$; these correspond to the data association constraints.
Furthermore, there are ``middle'' loops across different data source blocks; these correspond to the incorporation of the observations provided
by all the data sources $s \! \in \! \Set{S}_{0}$. Finally, there is an ``outer'' loop across all the PTs, which is due to the factor $\chi\big(\V{\tau}_{n},M_{n}^{(0)}\big)$.

\section{SPA-Based Data Fusion and \\ Multitarget Tracking}
\label{sec:BP_approach}

\begin{figure}[!t]

\vspace{1mm}

\centering
\includegraphics{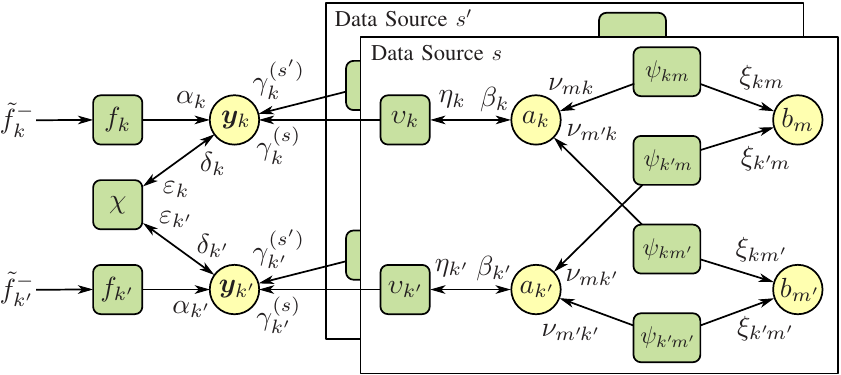}

\vspace{-.5mm}

\caption{Factor graph representing the factorization \eqref{eq:pdf_factorisation_final} of the joint posterior pdf $f(\V{y},\V{a},\V{b} | \V{z} )$ for one time step $n$.
For conciseness, the indices $n$, $j$ (inner-loop iteration index), and $i$ (outer-loop iteration index) are omitted, and the following short notations are used: 
$\tilde{f}^{-}_k \! \deq \! \tilde{f} (\V{y}_{n - 1,k})$,
$\V{y}_{k} \! \deq \! \V{y}_{n,k}$,
$a_{k} \! \deq \! a_{n,k}^{(s)}$,
$b_{m} \! \deq \! b_{n,m}^{(s)}$,
$f_k \! \deq \! f (\V{y}_{n,k} | \V{y}_{n-1,k})$,
$\chi \! \deq \! \chi\big(\V{\tau}_{n},M_{n}^{(0)}\big)$,
$\upsilon_{k} \! \deq$ $\upsilon^{(s)}\big(\V{y}_{n,k},a_{n,k}^{(s)};\V{z}_{n}^{(s)}\big)$,
$\psi_{km} \!\rmv \deq \! \psi^{(s)}\big(a_{n,k}^{(s)},b_{n,m}^{(s)}\big)$,
$\alpha_k \!\rmv \deq \! \alpha(\V{y}_{n,k})$,
$\delta_{k} \! \deq$ $\delta^{(i)}(\V{y}_{n,k})$,
$\varepsilon_{k} \! \deq \! \varepsilon^{(i)}(\V{y}_{n,k})$,
$\gamma_{k}^{(s)} \!\rmv \deq \! \gamma^{(s)(i)}(\V{y}_{n,k})$,
$\eta_{k} \!\rmv \deq \! \eta^{(s)(i)} \big( a_{n,k}^{(s)} \big)$,
$\beta_{k} \!\rmv \deq \! \beta^{(s)(i)} \big( a_{n,k}^{(s)} \big)$,
$\nu_{mk} \! \deq\rmv \nu_{m \rightarrow k}^{(s)(j)}\big(a_{n,k}^{(s)}\big)$, and 
$\xi_{km} \! \deq \xi_{k \rightarrow m}^{(s)(j)}\big(b_{n,m}^{(s)}\big)$.
}

\label{fig:factor_graph}

\vspace{-1mm}

\end{figure}

The posterior pdfs $f(\V{y}_{n,k} | \V{z}) \rmv=\rmv f (\V{x}_{n,k}, r_{n,k}, \tau_{n,k} | \V{z})$, $k \!\in\! \Set{K}$, used for
PT existence confirmation and PT state estimation as discussed in Section~\ref{sec:proposed_algorithm}, 
are marginal densities of the joint posterior pdf $f (\V{y},\V{a},\V{b} | \V{z})$. Direct marginalization is generally infeasible as it requires high-dimensional integrations and summations.
However, following \cite{MeyBraWilHla:J17}, approximations
of the marginal posterior pdfs $f(\V{y}_{n,k} | \V{z})$
can be calculated efficiently by applying the SPA \cite{KscFreLoe:01,Loe:04} to the factor graph in Fig.~\ref{fig:factor_graph}.
Because that factor graph contains loops, the SPA is executed iteratively. We will compute the individual SPA messages 
in an order that is defined by the following rules:
(i) Messages are not sent backward in time; (ii) \emph{iterative} message passing is
performed
within the inner loops and
the outer loop, and not for the middle loops.
As a consequence, the messages at
different data source blocks ($\eta_{k}$, $\beta_{k}$, $\nu_{mk}$, and $\xi_{km}$ in Fig.~\ref{fig:factor_graph}) can
be computed in parallel, without any direct interaction between them.

In the 
next subsections, we will
present expressions of the SPA messages that are passed between the various nodes of the factor graph in Fig.~\ref{fig:factor_graph},
as shown there. 
Some of these expressions were already provided
in \cite{gaglione18}.
The expressions are obtained by direct application of the general SPA rules described in \cite{KscFreLoe:01,Loe:04}, 
and are introduced
in what follows in
the order in which they are calculated at each time step $n$.
We will denote by $I$ and $i \! \in \! \{1, \ldots, I\}$ the number of iterations and the iteration index, respectively, for the outer loop, 
and by $J$ and $j \! \in \! \{1, \ldots, J\}$ the number of iterations and the iteration index, respectively, for the inner loop.
Furthermore, the final approximation of the marginal posterior pdf 
$f(\V{y}_{n,k} | \V{z})$\linebreak is denoted by $\tilde{f}(\V{y}_{n,k})$ and referred to as a belief.

\vspace{-1mm}

\subsection{Prediction}
\label{sec:pred}

In the prediction step, the beliefs computed at the previous time step, $\tilde{f}(\V{y}_{n-1,k}) = \tilde{f}(\V{x}_{n-1,k},r_{n-1,k},\tau_{n-1,k})$,
are propagated to the current time $n$ and converted into messages $\alpha(\V{y}_{n,k}) = \alpha(\V{x}_{n,k},r_{n,k},\tau_{n,k})$ according to
\begin{align}
	&\alpha(\V{x}_{n,k},r_{n,k},\tau_{n,k}) \nonumber \\[0mm]
	&=\! \sum_{r_{n-1,k} \in \{0,1\}} \sum_{\tau_{n-1,k} \in \Set{D}_{0}} \ist \int \rmv \tilde{f}(\V{x}_{n-1,k},r_{n-1,k},\tau_{n-1,k}) \nonumber \\[1mm]
	&\hspace{7mm}\times\rmv f(\V{x}_{n,k}, r_{n,k}, \tau_{n,k} | \V{x}_{n-1,k},r_{n-1,k}, \tau_{n-1,k}) \ist d\V{x}_{n-1,k} \nonumber \\[1mm]
	&=\! \sum_{r_{n-1,k} \in \{0,1\}} \sum_{\tau_{n-1,k} \in \Set{D}_{0}} \! p(\tau_{n,k} | r_{n,k},r_{n - 1,k}, \tau_{n-1,k})\nonumber \\
	&\hspace{7mm}\times\! \int \rmv f(\V{x}_{n,k}, r_{n,k} | \V{x}_{n-1,k},r_{n-1,k}) \nonumber \\[-1mm]
	&\hspace{16mm}\times\rmv \tilde{f}(\V{x}_{n-1,k},r_{n-1,k},\tau_{n-1,k}) \ist d\V{x}_{n-1,k} \ist,
	\label{eq:alpha_k}
\end{align}
where we used \eqref{eq:augmented_state_transition_pdf} for $f ( \V{x}_{n,k}, r_{n,k}, \tau_{n,k} | \V{x}_{n-1,k}, r_{n-1,k},\linebreak \tau_{n-1,k} )$.
As shown in Fig.~\ref{fig:factor_graph}, the messages $\alpha(\V{x}_{n,k},r_{n,k},\tau_{n,k})$
are passed from the factor nodes ``$f_{k}$'' to the variable nodes ``$\V{y}_{k}$''.
From \eqref{eq:alpha_k} and the fact that $\tilde{f}(\V{x}_{n-1,k},r_{n-1,k},\tau_{n-1,k})$ is normalized,
it follows that $\alpha(\V{x}_{n,k},r_{n,k},\tau_{n,k})$ is normalized too,
i.e.,
$ \sum_{r_{n,k} \in \{0,1\}} \rmv\sum_{\tau_{n,k} \in \Set{D}_{0}} \rmv\int \! \alpha(\V{x}_{n,k},r_{n,k},\tau_{n,k}) \ist d\V{x}_{n,k} \!=\!\rmv 1$. 
It will be convenient to introduce
\begin{align}
	\nonumber \\[-6mm]
	\alpha_{n,k} &\deq\! \sum_{\tau_{n,k} \in \Set{D}_{0}} \int \! \alpha(\V{x}_{n,k},0,\tau_{n,k}) \ist d\V{x}_{n,k}  \nonumber \\[.5mm]
	&= \int \! \alpha(\V{x}_{n,k},0,0) \ist d\V{x}_{n,k} \nonumber \\[-.5mm]
	&= 1 \rmv-\! \sum_{\tau_{n,k} \in \Set{D}_{0}} \int \! \alpha(\V{x}_{n,k},1,\tau_{n,k}) \ist d\V{x}_{n,k} \ist,
	\label{eq:alpha_k_0_integrated} \\[-6mm]
	\nonumber
\end{align}
where the second step follows from \eqref{eq:single_target_label_pdf_1}.
We note that 
$1 \!-\rmv \alpha_{n,k}$\linebreak $= \sum_{\tau_{n,k} \in \Set{D}_{0}}  \ist \int \rmv \alpha(\V{x}_{n,k},1,\tau_{n,k}) \ist d\V{x}_{n,k}$ 
can be interpreted as the predicted probability of existence of PT $k$.

\vspace{-1mm}

\subsection{Outer Loop}
\label{subsec:outer_loop}

The outer loop 
is composed of the factor node ``$\chi$'', the variable nodes ``$\V{y}_{k}$'', and the
blocks designated
``Data Source'' 
in Fig.~\ref{fig:factor_graph}; the latter process the observations provided by the TIPS sensors and the TDRS.
The messages passed from ``$\chi$'' to ``$\V{y}_{k}$'' in the $i$th outer-loop iteration are calculated as
\begin{align}
	\delta^{(i)}(\V{y}_{n,k}) &= \! \sum_{\V{\tau}_{n,k}^{-} \in \Set{D}_{0}^{K-1}} \sum_{\V{r}_{n,k}^{-} \in \{0,1\}^{K-1}} \!\!\chi\big(\V{\tau}_{n},M_{n}^{(0)}\big) \nonumber \\[0mm]
	&\hspace{7mm} \times\!\!\prod_{k' \in \Set{K} \setminus \{k\}} \int \rmv \varepsilon^{(i)}(\V{y}_{n,k'}) \ist d\V{x}_{n,k'} \nonumber \\[.5mm]
	&=\! \sum_{\V{\tau}_{n,k}^{-} \in \Set{D}_{0}^{K-1}} \!\!\chi\big(\V{\tau}_{n},M_{n}^{(0)}\big) \!\rmv\prod_{k' \in \Set{K} \setminus \{k\}} \!\! \tilde{\varepsilon}^{(i)}(\tau_{n,k'}) \ist. \nonumber \\[-4mm]
	\label{eq:delta_k} \\[-8mm]
	\nonumber
\end{align}
Here, $\V{r}_{n,k}^{-} \!\deq \! [r_{n,1},\ldots,r_{n,k-1},r_{n,k+1},\ldots,r_{n,K}]^{\T}$ is the vector of all the existence variables except the $k$th;
$\V{\tau}_{n,k}^{-} \! \deq \! [\tau_{n,1}, \linebreak \ldots,\tau_{n,k-1},\tau_{n,k+1},\ldots,\tau_{n,K}]^{\T}$ is the vector of all the
TID variables except the $k$th;
$\varepsilon^{(i)}(\V{y}_{n,k})$ are the messages passed from ``$\V{y}_{k}$'' to ``$\chi$'', which are calculated as
\begin{align}
	\nonumber \\[-6mm]
	\varepsilon^{(i)}(\V{y}_{n,k}) = \alpha(\V{y}_{n,k}) \prod_{s = 0}^{S} \gamma^{(s)(i)}(\V{y}_{n,k}) \ist;
	\label{eq:epsilon_k} \\[-8mm]
\nonumber
\end{align}
and
\begin{align}
	\tilde{\varepsilon}^{(i)}(\tau_{n,k}) \deq\! \sum_{r_{n,k} \in \{ 0,1 \}} \int \! \varepsilon^{(i)} (\V{x}_{n,k},r_{n,k},\tau_{n,k}) \ist d\V{x}_{n,k} \ist.
	\label{eq:epsilon_k_marginalised}
\end{align}
The messages $\gamma^{(s)(i)}(\V{y}_{n,k})$ in \eqref{eq:epsilon_k}, which are passed from the data source blocks to ``$\V{y}_{k}$'',
will be presented
in Section~\ref{subsec:data_association}.
We finally note that,
according to expression \eqref{eq:delta_k},
$\delta^{(i)}(\V{y}_{n,k})$ depends only on 
$\tau_{n,k}$. Thus, we will denote it as $\tilde{\delta}^{(i)}(\tau_{n,k})$. 

\vspace{-1mm}

\subsection{Observation Evaluation}
\label{sec:observ}

A further operation
within the $i$th outer-loop iteration is the calculation of
the messages
\begin{align}
	\beta^{(s)(i)} \big( a_{n,k}^{(s)} \big) &= \!\!\! \sum_{\tau_{n,k} \in \Set{D}_{0}} \!\!\! \tilde{\delta}^{(i-1)}(\tau_{n,k}) \!\!\!\!\! \sum_{r_{n,k} \in \{0,1\}} \rmv \int \! \alpha(\V{x}_{n,k},r_{n,k},\tau_{n,k}) \nonumber \\[.5mm]
	&\hspace{3.5mm} \times \upsilon^{(s)} \big( \V{x}_{n,k},r_{n,k},\tau_{n,k},a_{n,k}^{(s)}; \V{z}_{n}^{(s)} \big) \ist d\V{x}_{n,k} 
	\label{eq:beta_k}
\end{align}
for all PTs $k \! \in \! \Set{K}$ and data sources $s \! \in \! \Set{S}_{0}$, with $\tilde{\delta}^{(0)}(\tau_{n,k}) = 1$.
These messages are passed from the factor nodes ``$\upsilon_{k}$'' to the variable nodes ``$a_{k}$''.
Differently from the previously presented messages, $\beta^{(s)(i)} \big( a_{n,k}^{(s)} \big)$ involves the observation $\V{z}_{n}^{(s)}\rmv$.
Let us introduce the
shorthand $\beta^{(s)(i)}_{n,k}(m) \deq \beta^{(s)(i)} \big( a_{n,k}^{(s)} \!=\! m \big)$.
Then, for the TDRS, i.e., $s \! = \! 0$, using 
\eqref{eq:function_v} and the definitions of the functions $h^{(0)} \big( \V{x}_{n,k},r_{n,k}, \tau_{n,k}, a_{n,k}^{(0)}; M_{n}^{(0)} \big)\rmv$ and
$g^{(0)} \big( \V{x}_{n,k},\linebreak r_{n,k},\tau_{n,k}, a_{n,k}^{(0)}; \V{z}_{n}^{(0)} \big)$ introduced in Section~\ref{sec:joint} and Section~\ref{sec:global-likelihood}, respectively,
we obtain for all $m \! \in \! \Set{M}_{n}^{(0)}$
\begin{align}
	\beta^{(0)(i)}_{n,k}(m) &= \! \sum_{\tau_{n,k} \in \Set{D}} \! \tilde{\delta}^{(i - 1)}(\tau_{n,k}) \, p (\zeta_{n,m} | \tau_{n,k}) \nonumber\\[.5mm]
	&\hspace{5mm}\times\! \int \! f(\V{q}_{n,m} | \V{x}_{n,k}) \ist \alpha(\V{x}_{n,k},1,\tau_{n,k}) \ist d\V{x}_{n,k} \ist,
	\label{eq:beta_k_ais_m} \\[-7.5mm]
\nonumber
\end{align}
and for $m \!=\! 0$
\begin{align}
	\beta^{(0)(i)}_{n,k}(0) &= \! \sum_{\tau_{n,k} \in \Set{D}_{0}} \!\! \tilde{\delta}^{(i - 1)}(\tau_{n,k}) \nonumber\\[0mm]
	&\hspace{5mm}\times\! \Big( \alpha_{n,k} \ist \mathds{1}(\tau_{n,k}) + \! \int \! \alpha(\V{x}_{n,k},1,\tau_{n,k}) \ist d\V{x}_{n,k} \Big) \ist,
	\label{eq:beta_k_ais_0} 
\end{align}
where definition \eqref{eq:alpha_k_0_integrated} has been used.
For the TIPS sensors $s \! \in \! \Set{S}$, 
expression \eqref{eq:beta_k} 
specializes
as follows:
\begin{align}
	\beta^{(s)(i)} \big( a_{n,k}^{(s)} \big) &= \tilde{\delta}^{(i-1)}(\tau_{n,k} \!=\rmv 0) \ist  \alpha_{n,k} \ist \mathds{1}\big( a_{n,k}^{(s)} \big) \ist \nonumber \\[1mm]
	&\hspace{4mm} +\!\rmv \sum_{\tau_{n,k} \in  \Set{D}_{0}} \!\!\tilde{\delta}^{(i-1)}(\tau_{n,k}) \! \int \! \alpha(\V{x}_{n,k},1,\tau_{n,k}) \nonumber \\[.5mm]
	&\hspace{7.5mm} \times \upsilon^{(s)} \big( \V{x}_{n,k},1,\tau_{n,k},a_{n,k}^{(s)}; \V{z}_{n}^{(s)} \big) \ist d\V{x}_{n,k} \ist. 
	\label{eq:beta_k_radar}
\end{align}
The messages $\beta^{(s)(i)} \big( a_{n,k}^{(s)} \big)$ 
are used for the iterative probabilistic data association 
algorithm\footnote{We
note that in an RFS-based filter derivation, probabilistic data association can be interpreted as an approximation of multi-Bernoulli mixture components 
by multi-Bernoulli components \cite{Wil:15}.}
discussed next.

\vspace{-1mm}

\subsection{Inner Loop: Iterative Probabilistic Data Association}
\label{subsec:data_association}

Iterative probabilistic data association corresponds to passing messages $\nu_{m \rightarrow k}^{(s)(j)}\big(a_{n,k}^{(s)}\big)$ and 
$\xi_{k \rightarrow m}^{(s)(j)}\big(b_{n,m}^{(s)}\big)$ on the inner loop. 
In the $j$th inner-loop iteration, following \cite{WilLau:J14}, these messages are calculated for each PT $k \! \in \! \Set{K}$, data source $s \! \in \! \Set{S}_{0}$, 
and observation $m \! \in \! \Set{M}_{n}^{(s)}$ according to
\begin{align}
	\xi_{k \rightarrow m}^{(s)(j)}\big(b_{n,m}^{(s)}\big) &=\! \sum_{a_{n,k}^{(s)} = 0}^{M_{n}^{(s)}} \!\!\psi^{(s)}\big(a_{n,k}^{(s)},b_{n,m}^{(s)} \big) \ist \beta^{(s)(i)} \big( a_{n,k}^{(s)} \big) \nonumber \\[0mm]
	&\hspace{7mm} \times \!\!\prod_{m' \in \Set{M}_{n}^{(s)} \setminus \{m\}} \!\!\! \nu_{m' \rmv\rightarrow k}^{(s)(j-1)}\big(a_{n,k}^{(s)}\big)
	\label{eq:xi_full}
	\\[-8mm] \nonumber
\end{align}
and
\begin{align}
	\nu_{m \rightarrow k}^{(s)(j)}\big(a_{n,k}^{(s)}\big) &=\! \sum_{b_{n,m}^{(s)} = 0}^{K} \!\!\psi^{(s)}\big(a_{n,k}^{(s)},b_{n,m}^{(s)} \big) \!\rmv\prod_{k' \in \Set{K} \setminus \{k\}} \!\!\! \xi_{k' \!\rightarrow m}^{(s)(j)}\big(b_{n,m}^{(s)}\big) \ist.\nonumber \\[-4mm]
	\label{eq:nu_full} \\[-6mm]
\nonumber
\end{align}
(We note that these messages also depend on the outer-loop iteration index $i$, which is omitted to simplify the notation.) 
The iteration constituted by \eqref{eq:xi_full} and \eqref{eq:nu_full} is initialized by $\nu_{m \rightarrow k}^{(s)(0)}\big(a_{n,k}^{(s)}\big) \! = \! 1$.
After all the inner-loop
iterations $j = 1,\ldots,J$ have been performed, the messages 
$\nu_{m \rightarrow k}^{(s)(J)}\big(a_{n,k}^{(s)}\big)$ are available. These messages are then used to calculate messages
$\eta^{(s)(i)}\big(a_{n,k}^{(s)}\big)$, which are passed from variable nodes ``$a_{k}$'' back to factor nodes ``$\upsilon_{k}$'', according to
\begin{align}
	\nonumber \\[-5.5mm]
	\eta^{(s)(i)}\big(a_{n,k}^{(s)}\big) =\! \prod_{m \in \Set{M}_{n}^{(s)}} \!\!\rmv \nu_{m \rightarrow k}^{(s)(J)}\big(a_{n,k}^{(s)}\big) \ist.
	\label{eq:eta} \\[-5.5mm]
	\nonumber
\end{align}
Finally, messages $\gamma^{(s)(i)}(\V{y}_{n,k})$, which are passed from factor nodes ``$\upsilon_{k}$'' to variable nodes ``$\V{y}_{k}$'', are calculated 
\vspace{-.5mm}
as
\be
\hspace{.3mm}\gamma^{(s)(i)}(\V{y}_{n,k}) =\! \sum_{a_{n,k}^{(s)} = 0}^{M_{n}^{(s)}} \!\! \upsilon^{(s)}\big(\V{y}_{n,k},a_{n,k}^{(s)}; \V{z}_{n}^{(s)}\big) \ist \eta^{(s)(i)}\big(a_{n,k}^{(s)}\big) \ist. 	
\!\!\!\!\label{eq:update_message} 
\ee

Following \cite{WilLau:J14} with certain modifications, an efficient implementation of the above algorithm with a complexity of $O(K M_{n}^{(s)})$ 
per inner-loop iteration can be developed. This implementation is based on the fact that, 
due to the definition of the binary function $\psi^{(s)}\big(a_{n,k}^{(s)},b_{n,m}^{(s)} \big)$
in 
Section~\ref{subsec:target_observation_association}, the expressions \eqref{eq:xi_full} and \eqref{eq:nu_full} can take on only two different values.
A detailed derivation and formulation is provided in the supplementary material manuscript \cite{SupplMat}.

\noeqref{eq:eta,eq:update_message}

\vspace{-1mm}

\subsection{Belief Calculation}
\label{subsec:belief}

After all the outer-loop
iterations $i = 1, \ldots,I$ have been performed, the messages $\tilde{\delta}^{(I)}(\tau_{n,k})$ and $\gamma^{(s)(I)}(\V{y}_{n,k})$
are available. The final step is to calculate the beliefs $\tilde{f}(\V{x}_{n,k}, \linebreak r_{n,k},\tau_{n,k})$ approximating the marginal posterior pdfs $f(\V{x}_{n,k}, \linebreak r_{n,k},\tau_{n,k} | \V{z})$ as 
\vspace{-.5mm}
\begin{align}
	\hspace{-3mm}\tilde{f}(\V{x}_{n,k},r_{n,k},\tau_{n,k}) &= \frac{1}{C_{n,k}} \ist \alpha(\V{x}_{n,k},r_{n,k},\tau_{n,k}) \ist \tilde{\delta}^{(I)}(\tau_{n,k}) \nonumber \\
	&\hspace{5mm} \times \prod_{s = 0}^{S} \gamma^{(s)(I)}(\V{x}_{n,k},r_{n,k},\tau_{n,k})\ist, 
	\label{eq:belief} \\[-5.5mm]
	\nonumber
\end{align}
with
$C_{n,k} \rmv\deq\rmv \sum_{r_{n,k} \in \{0,1\}} \sum_{\tau_{n,k} \in \Set{D}_{0}} \int \rmv\tilde{f}(\V{x}_{n,k},r_{n,k},\tau_{n,k}) \ist d\V{x}_{n,k}$.

\noeqref{eq:belief}

\vspace{-1mm}

\subsection{Implementation and Complexity Reduction}
\label{sec:implementation_details}

\begin{algorithm}[!t]
\vspace{.5mm}
\
\caption{\small SPA-Based MTT Algorithm with TIPS-TDRS Fusion}
\label{alg:spa-based-mtt}

\fontsize{8}{10.5}\selectfont

\begin{algorithmic}[1]

\Require $\tilde{f} (\V{y}_{n-1,k})$, $\V{z}_{n}$.

\vspace{.5mm}

\Ensure $\tilde{f} (\V{y}_{n,k})$

\vspace{.9mm}

\State \textsc{Prediction} \, Compute $\alpha (\V{y}_{n,k})$ for all $k \! \in \! \Set{K}$ as in \eqref{eq:alpha_k}

\vspace{.5mm}
\Statex \textsc{Outer Loop}

\State $\delta^{(0)} (\V{y}_{n,k}) \longleftarrow 1$

\For{$i = 1$ \textbf{to} $I$}

\vspace{.5mm}
\Statex \textsc{Observation Evaluation}

\For{$s = 0$ \textbf{to} $S$} (in parallel)

\State Compute $\beta^{(s)(i)} (a_{n,k}^{(s)})$ for all $k \! \in \! \Set{K}$ as in \eqref{eq:beta_k}

\vspace{.3mm}

\State \textsc{Inner Loop} \, Compute data association messages \eqref{eq:xi_full} and \eqref{eq:nu_full}

\vspace{.3mm}

\State Compute $\eta^{(s)(i)} (a_{n,k}^{(s)})$ for all $k \! \in \! \Set{K}$ as in \eqref{eq:eta}

\State Compute $\gamma^{(s)(i)} (\V{y}_{n,k})$ for all $k \! \in \! \Set{K}$ as in \eqref{eq:update_message}

\EndFor

\vspace{.5mm}

\State Compute $\varepsilon^{(i)} (\V{y}_{n,k})$ for all $k \! \in \! \Set{K}$ as in \eqref{eq:epsilon_k}

\State Compute $\delta^{(i)} (\V{y}_{n,k})$ for all $k \! \in \! \Set{K}$ as in \eqref{eq:delta_k}

\vspace{.5mm}

\EndFor

\vspace{.5mm}

\State \textsc{Belief Computation} \, Compute $\tilde{f} (\V{y}_{n,k})$ for all $k \! \in \! \Set{K}$ as in \eqref{eq:belief}

\vspace{1mm}

\end{algorithmic}

\vspace{-1mm}

\end{algorithm}

A step-by-step summary of the proposed SPA-based MTT algorithm with TIPS-TDRS fusion is provided in
Algorithm~\ref{alg:spa-based-mtt}.
A particle-based implementation can be obtained 
by extending the algorithm presented in \cite{MeyBraWilHla:J17}. This implementation approximates the beliefs $\tilde{f}(\V{y}_{n,k})$ as well as the messages $\alpha(\V{y}_{n,k})$ and 
$\gamma^{(s)(i)}(\V{y}_{n,k})$ by a set of particles and corresponding weights \cite{AruMasGorCla:02}, 
thereby avoiding the direct computation of integrals and enabling a feasible computation of the messages and beliefs.

However, the computation of the message $\tilde{\delta}^{(i)}(\tau_{n,k})$ can still be expensive;
indeed, according to \eqref{eq:delta_k}, 
$\tilde{\delta}^{(i)}(\tau_{n,k})$ is the sum of $(D + 1)^{K - 1}$
terms, and thus its 
computation scales exponentially in the number $K$ of PTs.
We therefore propose a low-complexity (LC) computation in which $\tilde{\delta}^{(i)} (\tau_{n,k})$ is approximated with only a single term, given by
\be
	\tilde{\delta}^{(i)}(\tau_{n,k}) \approx \chi\big(\V{\tau}_{n}^{\star},M_{n}^{(0)}\big) \!\rmv\prod_{k' \in \Set{K} \setminus \{k\}} \!\! \tilde{\varepsilon}^{(i)}(\tau_{n,k'}^{\star}) \ist,
	\label{eq:delta_k_lc}
\vspace{-1mm}
\ee
where $\V{\tau}_{n}^{\star} \deq [ \tau_{n,1}^{\star},\ldots,\tau_{n,k-1}^{\star},\tau_{n,k},\tau_{n,k+1}^{\star},\ldots,\tau_{n,K}^{\star} ]^{\T}$ and $\tau_{n,k}^{\star} \deq \argmax_{\,\tau_{n,k} \in \Set{D}_{0}} \tilde{\varepsilon}^{(i)}(\tau_{n,k})$.
To motivate
\eqref{eq:delta_k_lc}, we note that
at each step of the outer loop, 
$\tilde{\varepsilon}^{(i)}(\tau_{n,k})$ can be interpreted as the (nonnormalized)
probability distribution of $\tau_{n,k}$, the TID of PT $k$, after observation evaluation and data association.
The messages $\tilde{\varepsilon}^{(i)}(\tau_{n,k'})$ coming from all 
PTs $k' \!\not=\rmv k$ 
are then used to obtain $\tilde{\delta}^{(i)}(\tau_{n,k})$. More specifically,
according to the summation in \eqref{eq:delta_k},
all the possible combinations between the TIDs in $\Set{D}_{0}$ and the PTs (except the $k$th) are evaluated and
weighted by $\chi(\V{\tau}_{n},M_{n}^{(0)})$, and the corresponding terms are marginalized out.
The LC approximation \eqref{eq:delta_k_lc} 
then corresponds to considering only the most likely of
these combinations. In Section~\ref{sec:results_simulations}, we will verify experimentally
the validity of this approximation.

\section{Results for Simulated Data}
\label{sec:results_simulations}

We evaluate the performance of the proposed MTT algorithm with TIPS-TDRS fusion
in
a simulated scenario, and 
we compare it
with that
of three alternative algorithms.

\vspace{-1mm}

\subsection{Experiment Setup}
\label{sec:results_simulations_setup}

\begin{figure}[!b]

\vspace{-1.5mm}

\centering

\includegraphics{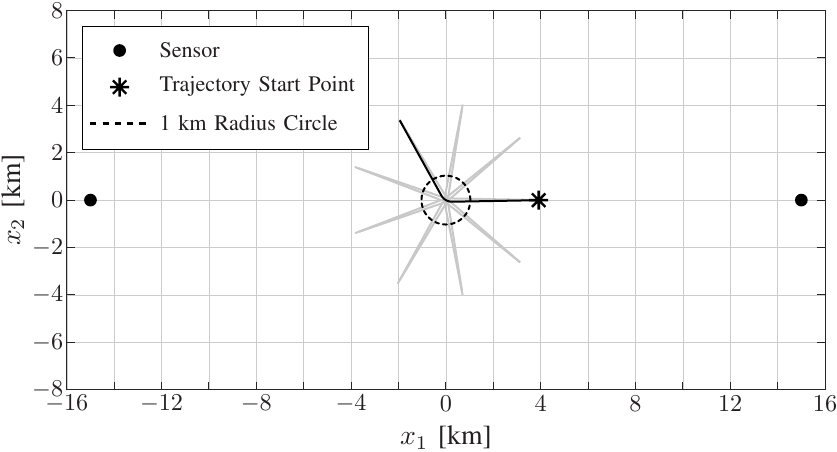}

\vspace{-1mm}

\caption{Simulated scenario: TIPS sensors (black bullets) and target trajectories. The star marks the initial position of the highlighted trajectory; 
the other trajectories are rotated (by $40$ degrees) versions
of the former.}

\label{fig:simulated-scenario}

\end{figure}

The simulated scenario consists of nine targets that are moving in a region of interest (ROI) with a constant velocity of $4$ m/s during
200 time steps, with time step duration $T \! = \! 10 \, \text{s}$.\linebreak The target trajectories, 
whose starting points are equally spaced on a circle with center 
$(0,0)$ and radius $4$ km, and
the ROI are shown
in Fig.~\ref{fig:simulated-scenario}. Five targets
appear at $n \! = \! 1$ and disappear at $n \! = \! 200$, and
the other
four targets appear at $n \! = \! 5$ and disappear at $n \! = \! 195$.
Six 
randomly selected targets 
are cooperative and transmit TDRS reports between $n \! = \! 10$ and $n \! = \! 190$; the number of TDRS reports transmitted per time interval is a random variable 
that is Poisson distributed with mean $0.5$ for three 
of the six cooperative targets and $2$ for the other
three cooperative targets.

The PT states $\V{x}_{n,k}$ are composed of 
the
2D position vector $\check{\V{x}}_{n,k}$ and the
2D velocity vector $\dot{\check{\V{x}}}_{n,k}$, 
i.e., $\V{x}_{n,k} \!= \! [\check{\V{x}}_{n,k}^{\T},\dot{\check{\V{x}}}_{n,k}^{\T}]^{\T} \!$.
For the dynamic model in \eqref{eq:dynamic_model1}, we choose the nearly constant velocity model, i.e., $\V{x}_{n,k} \! = \! \V{F} \V{x}_{n-1,k} + \V{G}\V{u}_{n,k}$,
where $\V{F} \! \in \! \RSet^{4 \times 4}$ and $\V{G} \! \in \! \RSet^{4 \times 2}$ are defined as in \cite[Sec. 6.3.2]{BarRonKir:01} (this involves
the time step duration
$T$) and
$\V{u}_{n,k} \! \sim \! \Set{N}(\V{0},\sigma_{\text{u}}^{2} {\bf I}_{2})$
is a 2D zero-mean iid Gaussian random vector that is also iid across $n$ and $k$, with $\sigma_{\text{u}} \! = \! 0.05$ m/s$^2$.

There
are two TIPS sensors (i.e., $S \! = \! 2$), which generate range-bearing measurements
$\V{z}_{n,m}^{(s)} \! = \! \big[ z_{n,m,\text{r}}^{(s)} \ist, z_{n,m,\text{b}}^{(s)} \big]^{\T}\rmv$, $s \! \in \! \Set{S}$. \linebreak
The range measurement $z_{n,m,\text{r}}^{(s)}$ is a Gaussian random variable with mean $\norm{\check{\V{x}}_{n,k} \!-\rmv \V{\rho}^{(s)}}$ and standard deviation 
$\sigma_{\text{r}}^{(s)} \! = \! \sigma_{\text{r}} \! = \! 250$ m, 
and
the bearing measurement $z_{n,m,\text{b}}^{(s)}$ is a von Mises random variable with mean $\angle ( \check{\V{x}}_{n,k} - \V{\rho}^{(s)} )$ and concentration parameter 
$\kappa_{\text{b}}^{(s)} \! = \! \kappa_{\text{b}} = 500$ \cite[Ch. 3.5.4]{MarJup:B09}.
Here,
$\V{\rho}^{(1)} \! = \! [15,0]^{\T}$ km and $\V{\rho}^{(2)} \! = \! [-15,0]^{\T}$ km
are the
positions
of 
the TIPS sensors.
Furthermore, the mean number of false alarms is $\mu^{(s)} \! = \! 2$ and the false alarm pdf $f_{\text{FA}}(\V{z}_{n,m}^{(s)})$ is uniform
on the ROI and zero outside.
The detection probability is $P_{\text{d}}^{(s)}(\V{x}_{n,k}) \! = \! P_{\text{d}}^{(s)} \! = \! 0.5$.

The TDRS self-measurement
is modeled as
$\V{q}_{n,m}^{(\ell)} \! = \rmv \check{\V{x}}_{n,k,m}^{(\ell)} \rmv + \V{v}_{n,m}^{(\ell)}$,
where 
$\V{v}_{n,m}^{(\ell)} \! \sim \! \Set{N} (\V{0}, \sigma_{\text{v}}^{2}{\bf I}_{2})$ with $\sigma_{\text{v}} \rmv=\rmv 10\,$m is a 2D zero-mean iid Gaussian random vector
that is also iid across $n$, $m$, and $\ell$.
The 
ID set is chosen as $\Set{D} \! = \! \{ 1,2,\ldots,6 \}$.
The probability that the
PT-to-TDRS transmission is successful and without errors is $p_{n,m}^{\text{c}} \!\rmv = \! p^{\text{c}} \! = \! 0.95$, 
and the probability that 
there is an error resulting in an RID outside $\Set{D}$
is $p_{n,m}^{\text{e}} \!\rmv = \! p^{\text{e}} \! = \! 0.045$; the RIDs within the reports are simulated in accordance with
these probabilities.
Finally, each TDRS report time $t_{n,m}^{(\ell)}$ is randomly (uniformly) chosen within the respective time step interval $(t_{n-1},t_{n}]$.

We compare our algorithm, which jointly processes TIPS and TDRS observations at each time step $n$, with three alternative algorithms.
The first conforms to the system model proposed in Section~\ref{sec:system_model_statistical_formulation}
except that the TDRS reports are sequentially processed as soon as they become available and, thus, the PT states are estimated each time a TDRS report is 
received; this will thus be called the ``sequential'' algorithm. The second alternative algorithm,
which we will refer to as the ``all-TIPS'' algorithm, models and treats the TDRS as a TIPS sensor.
Accordingly, the TDRS clusters are considered as TIPS measurements, which means that their RID is ignored (once the clusters are formed) and the possibility that they are false alarms is not ruled out.
For this virtual TIPS sensor, we assume a detection probability of $P_{\text{d}}^{(0)} \! = \! 0.9$,
$0.5$, or $0.25$---depending on the scenario---and 
a mean number of false alarms of $\mu^{(0)} \! = \! 10^{-16}$.
Here, the small value of
$\mu^{(0)}$ reflects the low mean number of false alarms expected from the TDRS. (We do not set $\mu^{(0)}$ to $0$, because this would result 
in a division by $0$ in 
the expression of $h^{(s)}(\cdot)$ reported below equation~\eqref{eq:prior_da_radar_factorisation_final}.) 
The third alternative algorithm is
the
one proposed in \cite{MeyBraWilHla:J17}, which fuses only the TIPS measurements 
and will hence be called the ``no-TDRS'' algorithm.
We use the following parameters in
all algorithms
if applicable and unless otherwise stated: 
$K = 30$, 
$p^{\text{t}}_{n,k} \! = \! p^{\text{t}} \! = \! 0.1$,
$p^{\text{b}}_{n,k} \! = \! p^{\text{b}} \! = \! 10^{-3}$,
$p^{\text{s}}_{n,k} \! = \! p^{\text{s}} \! = \! 0.999$,
and $P_{\text{th}} = 0.95$.

We assess the performance of the various algorithms in terms of the Euclidean distance based
generalized optimal sub-pattern assignment
error
for trajectories (GOSPA-T) \cite{GarRahAbuSve:J20},
with cut-off parameter $c \rmv=\rmv 500\,$m, switching penalty 
250 m, and order 1.
The GOSPA-T metric accounts for localization errors for correctly confirmed targets, 
errors for missed targets and false targets (i.e., confirmed PTs not corresponding to any actual target), and an error for track switches.
Furthermore, we report the time on target (ToT), 
track fragmentation (TF), and 
false alarm rate (FAR).
The ToT is the
fraction of time
during which the PT corresponding to a target is
confirmed and the distance between its estimated and true
positions is lower than
500 m.
The TF is the number of different PTs associated
with a target 
during the target's lifespan.
The FAR is the number of false tracks
per unit of space and unit of time.
As the ToT and TF are defined for each target individually, we will usually consider their averages taken over all 
targets, which will be referred to as A-ToT and A-TF, respectively.
Finally, to evaluate the accuracy in estimating the TIDs for the proposed algorithm and the sequential algorithm, 
we report a metric referred to as the TID errors count. At each time step $n$, given the optimal sub-pattern assignment (with cut-off parameter $c \rmv=\rmv 500\,$m) 
between the actual targets and the estimated PTs, the TID errors count is defined as the number of errors committed by the tracking algorithm in estimating all the TIDs.
All the mentioned performance indices are averaged over 100 simulation runs.

\vspace{-1mm}

\subsection{Effect of the LC Approximation}
\label{sec:first-scenario:effect}

\begin{figure}[!t]

\centering

\includegraphics{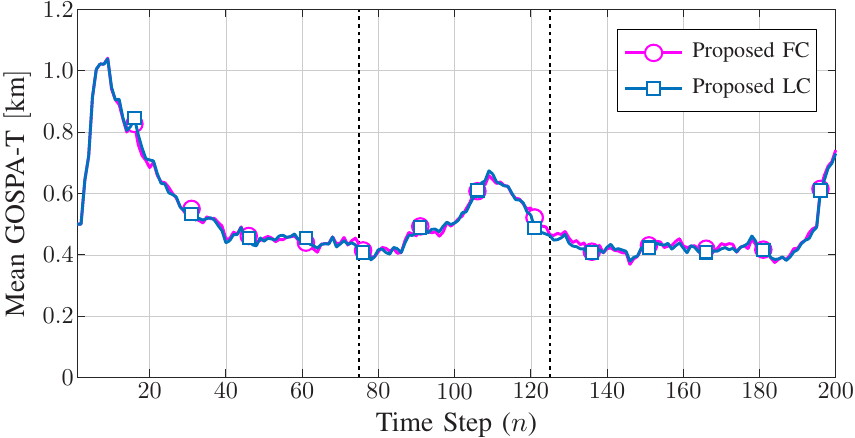}

\vspace{-1mm}

\caption{
Mean GOSPA-T obtained with the proposed algorithm using the FC and LC implementations of the outer loop.
In this figure and in subsequent figures, the dashed vertical lines delimit
the time interval during which the targets are within the circle of radius 1 km 
shown in Fig.\ \ref{fig:simulated-scenario}.}

\label{fig:LC-approx-MOSPA}

\vspace{-1mm}

\end{figure}

\begin{figure}[!t]

\centering

\includegraphics{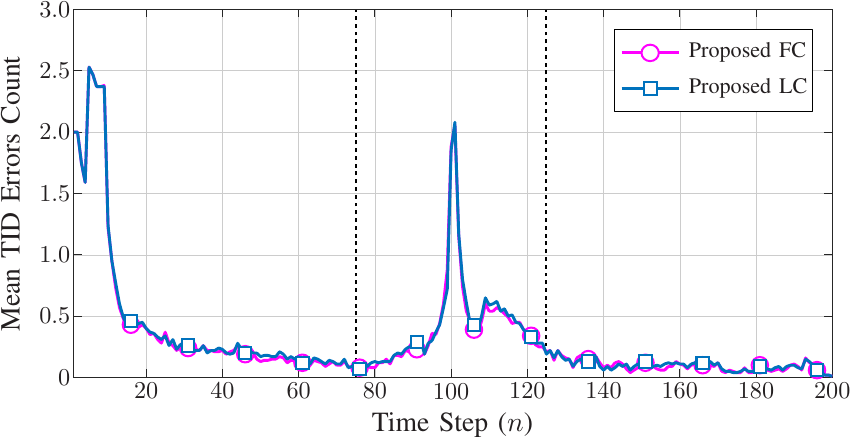}

\vspace{-1mm}

\caption{Mean TID errors count obtained with the proposed algorithm using the FC and LC implementations of the outer loop.}

\label{fig:LC-approx-TID-errors-count}

\vspace{-1mm}

\end{figure}

\begin{figure}[!t]

\centering

\includegraphics{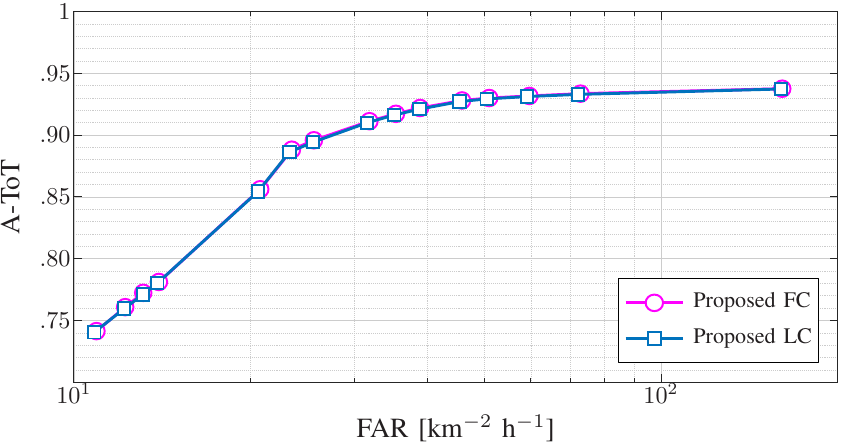}

\vspace{-.5mm}

\caption{A-ToT versus FAR obtained for varying
existence confirmation threshold $P_{\text{th}}$ using the FC and LC implementations of the outer loop.}

\label{fig:LC-approx-far-tot}

\vspace{-1mm}

\end{figure}

We first assess the validity of the LC approximation 
\eqref{eq:delta_k_lc} by comparing the resulting performance
with that of the full-complexity (FC) computation of the messages $\tilde{\delta}^{(i)}(\tau_{n,k})$ according to \eqref{eq:delta_k}.
To avoid an excessive complexity of the FC computation, we temporarily consider only
three targets---two cooperative and one noncooperative---and 
choose $K \!=\! 10$ and $\Set{D} \! = \! \{ 1,2 \}$. Fig.~\ref{fig:LC-approx-MOSPA} shows the
mean GOSPA-T of the LC and FC versions versus time step $n$.
One can see that the LC approximation does not reduce the performance of the proposed algorithm.
This is confirmed by Fig.~\ref{fig:LC-approx-TID-errors-count}, which shows the mean TID errors count, and by Fig.~\ref{fig:LC-approx-far-tot}, which
shows the A-ToT versus FAR performance for varying
existence confirmation threshold $P_{\text{th}}$ (cf.\ Section~\ref{sec:proposed_algorithm}).
A further confirmation is provided by the fact that the A-TF was obtained as $1.34$ for the FC version and $1.35$ for the LC version.


\subsection{Comparison with Alternative Algorithms}
\label{sec:first-scenario:comp}

\begin{figure}[!t]

\centering

\includegraphics{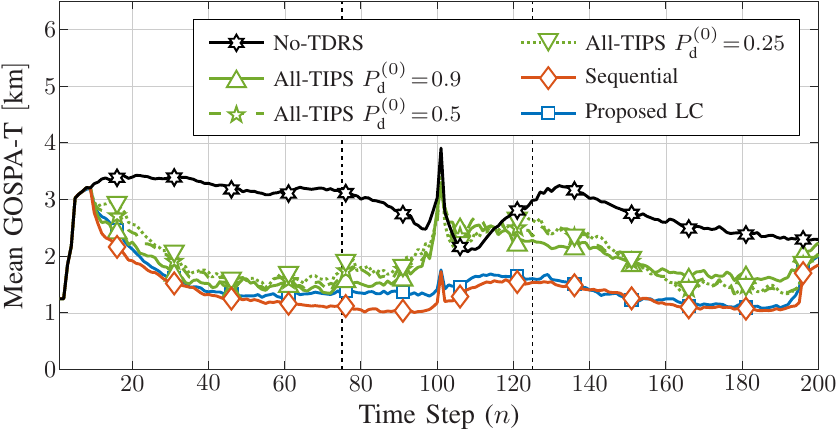}

\vspace{-1mm}

\caption{
Mean GOSPA-T obtained with the proposed algorithm (with LC approximation)
and with the sequential, 
all-TIPS, and 
no-TDRS algorithms.}

\label{fig:comparison-prop-alltips-notdrs}

\vspace{-1mm}

\end{figure}

Fig.~\ref{fig:comparison-prop-alltips-notdrs} compares the
mean GOSPA-T
of the proposed algorithm (with LC approximation), 
of the all-TIPS algorithm
with $P_{\text{d}}^{(0)} \!\! = \! 0.9$,
$P_{\text{d}}^{(0)} \!\! = \! 0.5$, and $P_{\text{d}}^{(0)} \!\! = \! 0.25$, of the sequential algorithm, and of the no-TDRS algorithm.
The proposed algorithm is seen to consistently outperform both the all-TIPS algorithm, independently of the 
chosen detection probability $P_{\text{d}}^{(0)}\rmv$, and the no-TDRS algorithm.
The performance difference between the proposed algorithm and the all-TIPS algorithm is largest in the time interval 
between time steps $n \!=\! 75$ and $n \!=\! 125$, when the targets are within the circle of radius 1~km shown in Fig.~\ref{fig:simulated-scenario}
and perform a smooth right turn.
Indeed, the exploitation of the RID included in the TDRS reports enables
the proposed algorithm to accurately track the targets even while maneuvering.
The performance of the proposed algorithm and the sequential algorithm is quite similar,
with a slightly
higher
mean GOSPA-T obtained with the proposed algorithm.
This is arguably due to the immediate processing of the TDRS reports performed by the sequential algorithm: because the TDRS reports are processed
as soon as they are received, the PT state prediction is performed over a shorter time interval, which
reduces the growth of uncertainty over time.
On the other hand, as shown next, the sequential processing leads to an inaccurate estimation of the TIDs.

\begin{figure}[!t]

\centering

\includegraphics{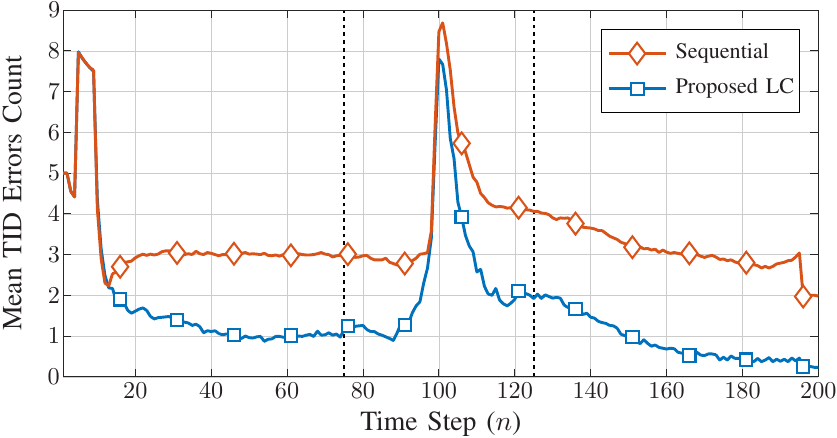}

\vspace{-1mm}

\caption{Mean TID errors count obtained with the proposed algorithm (with LC approximation) and the sequential algorithm.}

\vspace{-1mm}

\label{fig:comparison-prop-sequential}

\end{figure}

Fig.~\ref{fig:comparison-prop-sequential} presents a comparison
between the proposed algorithm and the sequential algorithm in terms of the mean TID errors count. 
It can be seen that the mean TID errors count of the sequential algorithm is consistently higher than that of the proposed algorithm, which indicates frequent errors in estimating the TIDs.
At time $n \!=\! 1$, the mean TID errors count is equal to 5 because only five targets are present in the ROI; it then increases at time $n \!=\! 5$ when the other four targets appear.
The peak at time $n \!=\! 100$ is mostly due to an erroneous assignment between the PTs and the actual targets,
which is
due to the close proximity of all the targets.

\begin{figure}[!t]

\centering

\includegraphics{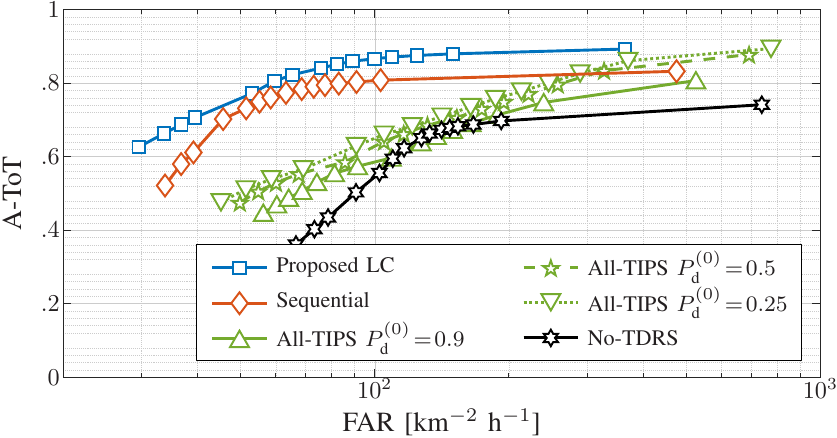}

\vspace{-1mm}

\caption{A-ToT versus FAR performance of 
the proposed algorithm (with LC approximation)
and 
of the sequential, 
all-TIPS, and 
no-TDRS algorithms.}

\label{fig:comparison-prop-all-far-tot}

\vspace{-1mm}

\end{figure}

Fig.~\ref{fig:comparison-prop-all-far-tot} compares the A-ToT versus FAR performance of the proposed algorithm and the three alternative algorithms.
It is seen that the proposed algorithm outperforms the other algorithms in that it yields the highest A-ToT for a given FAR.

Finally, the proposed algorithm outperforms the other algorithms also in terms of the A-TF. Indeed, we obtained its A-TF as $2.45$,
which is lower than the A-TF of the all-TIPS algorithm ($4.81$, $3.93$, and $3.67$ for $P_{\text{d}}^{(0)} \! = \! 0.9$, $0.5$, and $0.25$, respectively),
of the sequential algorithm ($2.62$), and of the no-TDRS algorithm ($3.64$).

\section{Results for Real Data}
\label{sec:results_realdata}

Next, we
apply the proposed algorithm to a real dataset acquired in a maritime
scenario and compare its
results with those obtained with the sequential algorithm.
The dataset has a duration of about 7.5 hours and consists of TIPS measurements provided by two 
radar sensors (i.e., $S \! = \! 2$) and TDRS reports provided by the AIS.
The radar measurements were acquired by two high-frequency surface wave 
(HFSW)
radars \cite{GurgelAES99_37}
located on the Italian coast, one on the island of Palmaria (IP) near La Spezia 
and the other in San Rossore Park (SRP) near Pisa.
Each measurement consists of range, bearing, and range rate,
i.e., $\V{z}_{n,m}^{(s)} \! = \! \big[ z_{n,m,\text{r}}^{(s)},z_{n,m,\text{b}}^{(s)},z_{n,m,\dot{\text{r}}}^{(s)} \big]^{\T}\rmv$.
The range measurement $z_{n,m,\text{r}}^{(s)}$ and bearing measurement $z_{n,m,\text{b}}^{(s)}$ are modeled as
described in Section~\ref{sec:results_simulations_setup}, with $\sigma_{\text{r}} \! = \! 150$ m and $\kappa_{\text{b}} \! = \! 1000$.
The range rate measurement is modeled as a Gaussian random variable with mean $(\check{\V{x}}_{n,k} \rmv-\rmv \V{\rho}^{(s)})^{\T} \ist \dot{\check{\V{x}}}_{n,k} / \norm{\check{\V{x}}_{n,k} \!-\rmv \V{\rho}^{(s)}}$  and standard deviation $\sigma_{\dot{\text{r}}}^{(s)} \!\rmv = \! \sigma_{\dot{\text{r}}} \! = \! 0.1$ m$/$s.
The mean number of false alarms is set to $\mu^{(s)} \!=\! 15$, and the false alarm pdf $f_{\text{FA}} (\V{z}_{n,m}^{(s)})$ is
uniform on the
ROI 
(which is
the intersection of the fields of view of the two radar sensors) and zero outside, 
as well as
uniform in range rate in the interval $[-25,25]$ m$/$s and zero outside.
The detection probability is set to $P_{\text{d}}^{(s)} \!=\rmv 0.8$.

Each AIS report contains
an RID, i.e., the maritime mobile service identity (MMSI), and 
a self-measurement of the position of the respective target (ship).
The self-measurement is
modeled as in Section~\ref{sec:results_simulations_setup}, with 
$\sigma_{\text{v}} \! = \! 100$ m.
Due to the lack of other information sources, the AIS data are also used as ground truth 
for evaluating
the time-averaged
mean GOSPA-T and the time-averaged mean TID errors count, as well as the A-ToT, A-TF, and FAR. 
More specifically, a sequence of reports with the same MMSI forms
an ``AIS track,'' 
which is considered to be a ground-truth trajectory.
Since these AIS tracks represent 
only a subset of the ships actually present in the ROI, our
performance assessment and comparison of the proposed algorithm and the sequential algorithm
are not exhaustive; 
they are merely intended to demonstrate the applicability of the proposed statistical formulation and tracking algorithms to a real-world scenario.
To account for the facts that the AIS sequences are not temporally aligned with the radar time steps
and some of them
contain ``gaps'' of several minutes or even hours,
we use cubic interpolation to determine the instantaneous positions
of the AIS tracks
at each radar time step.

The number of PTs is
set to $K \!=\! 100$.
The PT state and
dynamic model are defined as in Section~\ref{sec:results_simulations_setup}, with
$\sigma_{\text{u}} \! = \! 0.05$ m/s$^2$ and
$T \! = \!16.64$ s.
The
existence confirmation threshold is set to $P_{\text{th}} \!=\rmv 0.99$.
All the other parameters
involved in our algorithm are chosen as specified in Section~\ref{sec:results_simulations_setup}.

\begin{figure}[!t]

\centering

\includegraphics{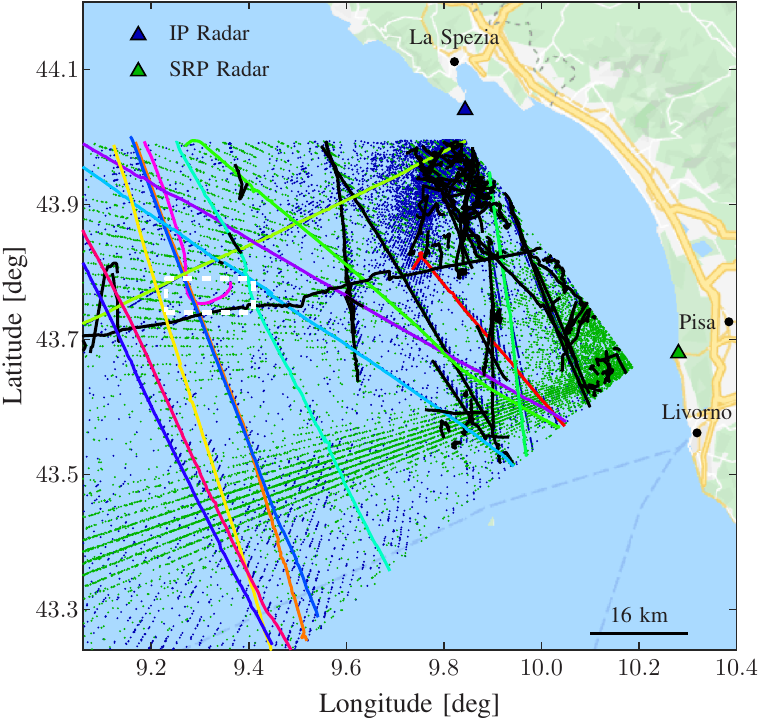}

\vspace{-1mm}
\caption{Measurements produced by the IP and SRP radar sensors 
(represented
by blue and green dots, respectively) and trajectories estimated 
by the proposed 
algorithm (represented
by black or colored lines) during 7.5 hours. 
Colored lines represent the estimated trajectories of
cooperative ships identified
through the MMSI, whereas black lines represent the estimated trajectories 
of noncooperative ships.
For improved clarity, trajectories of duration less than ten
time steps are not shown.
The dashed white rectangle
identifies the subregion
depicted in Fig.~\ref{fig:real_data_zoom}.}
\label{fig:HFSW_results}

\end{figure}

In Fig.~\ref{fig:HFSW_results}, we depict the measurements of the two radar sensors and the trajectories estimated by the proposed
algorithm during 7.5 hours. 
(The trajectories estimated by the sequential algorithm are very similar; they are not shown to avoid a cluttered figure.)
One can see that fusing the TIPS measurements provided by the HFSW radars and the TDRS reports provided by the AIS allows 
MMSI-based identification of the
cooperative ships.
\begin{figure}[!t]
	\centering
	
	\subfloat[Proposed algorithm.]{
\includegraphics{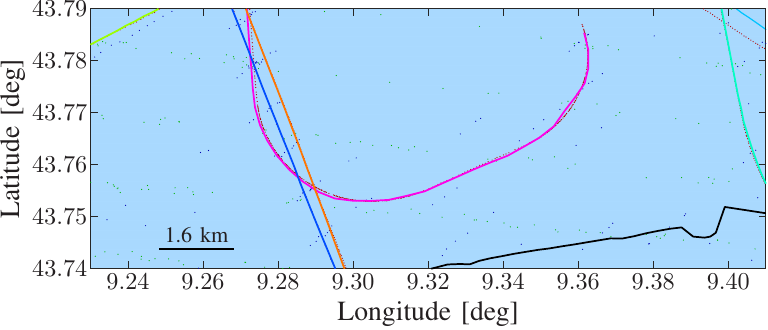}
		\label{fig:zoom_proposed} 
	}  
	\vspace{-2mm}
	\centering

	\subfloat[Sequential algorithm.]{
\includegraphics{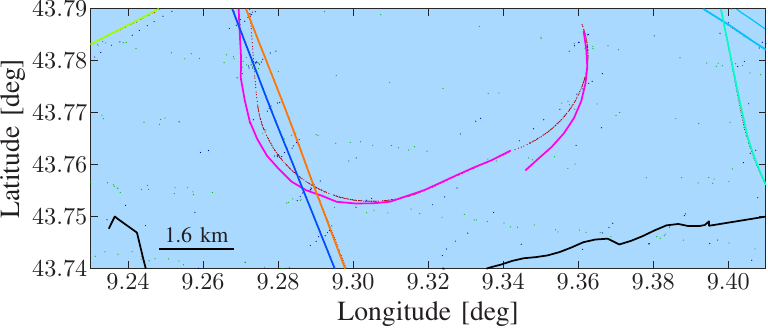}
		\label{fig:zoom_sequential} 
	}
	
	\caption{Trajectories estimated by (a) the proposed 
		algorithm and (b) the sequential algorithm
		in a 	rectangular subregion
		of the ROI.
		Blue and green dots represent
		measurements produced by the IP and SRP radar sensors, respectively, and 
		red dots represent
		the 
		positions indicated by
		the AIS reports.}
	
	\vspace{-2mm}
	
	\label{fig:real_data_zoom}
\end{figure}
On the other hand, the lack of AIS reports for the remaining noncooperative
ships does not impede their tracking.
Note that the superposition of the measurements seems to form line segments along some range directions of the radar sensors; these line segments are caused by a nonuniform distribution of false detections affecting HFSW radars, and do not correspond to actual ship trajectories.

Fig.~\ref{fig:real_data_zoom} compares the trajectories estimated by the proposed 
algorithm and the sequential algorithm
in a rectangular subregion
of the ROI where one of the ships
is maneuvering. We observe that in this specific example the proposed algorithm 
produces a single track for the maneuvering ship, whereas the sequential algorithm does not achieve track continuity.

The time-averaged mean GOSPA-T is $2353$ m for the proposed algorithm and $2042$ m for the sequential algorithm.
These results conform to those we obtained for simulated data in Section \ref{sec:first-scenario:comp}.
The time-averaged mean TID errors count is $0.05$ for the proposed algorithm and $0.78$ for the sequential algorithm, which confirms that the proposed algorithm 
is more accurate in estimating the TIDs. Furthermore, compared to the sequential algorithm, the proposed algorithm exhibits a higher A-ToT 
($0.99$ versus $0.82$), a higher FAR ($0.14$ 
versus $0.12$ km$^{-1}$h$^{-1}$),
and a higher A-TF ($1.31$ versus $1.23$).
Note that definite conclusions cannot be drawn from these results because the ground truth is available only for a subset of the ships, i.e., for those providing AIS reports.
Indeed, the results are mainly intended to demonstrate the applicability of the proposed statistical formulation and tracking algorithms to a real-world scenario.

\vspace{-.5mm}

\section{Conclusion}
\label{sec:conclusion}

Heterogeneous data fusion is an important functionality in high-performance multitarget tracking (MTT) systems. In this paper, we considered multisensor 
MTT with an inherent probabilistic fusion of two classes of data sources: sensors that produce measurements without requiring target cooperation, and
a reporting system that conveys information---possibly including target
ID---that is provided by cooperative targets. 
We established a statistical observation model that combines these two classes of data sources
and accounts for measurement-origin uncertainty, missed detections, false alarms, incorrectly received IDs, and asynchronicity.

Adopting a Bayesian framework, target
existence confirmation and state estimation essentially amount to marginalizing a joint posterior
distribution that involves the target states, 
existence indicators, and
IDs,
the
observation-target association indices, and
the past and current observations from
all the sensors and
the reporting system.
To obtain an efficient and scalable sequential algorithm for approximate marginalization that 
exploits conditional independencies, we used the sum-product algorithm on a factor graph that represents the 
structure of our Bayesian statistical model for the multisensor MTT and information fusion problem.
The resulting
MTT algorithm 
allows
a real-time integration of the two classes of data sources and exhibits high tracking accuracy at moderate complexity.
We demonstrated the performance of the
algorithm and the effectiveness of our fusion approach using both simulated data and real data 
from a maritime surveillance experiment. 
Our 
results showed the benefits of fusing heterogeneous information and
substantial performance advantages over
three alternative algorithms.

A possible direction of future research is the extension of the proposed framework to an indoor simultaneous localization and mapping scenario \cite{LeiMeyHlaWitTufWin:J19} enhanced by radio-frequency identification tags \cite{GueFraWilKosMilBryStrKle:C06}.

\ifCLASSOPTIONcaptionsoff
  \newpage
\fi

\balance
\renewcommand{\baselinestretch}{1}
\selectfont

\bibliographystyle{IEEEtran}
\bibliography{IEEEabrv,biblio}

\end{document}


\onehalfspacing

\title{Fusion of Sensor Measurements and Target-Provided Information\\[-1mm] in Multitarget Tracking --- Supplementary Material\vspace{2mm}}

\author{\hspace{-2.5mm}Domenico~Gaglione, Paolo~Braca, Giovanni~Soldi, Florian~Meyer, Franz~Hlawatsch, and Moe~Z.~Win\vspace{2mm}
}\date{\today\vspace{0mm}}

\maketitle

\renewcommand{\baselinestretch}{1.13}\small\normalsize

\noindent This\blfootnote{
This work was supported in part by the NATO Allied Command Transformation (ACT) under the DKOE project, 
by the Austrian Science Fund (FWF) under
grant P 32055-N31,
by the Czech Science Foundation (GA\v{C}R) under grant 17-19638S,
by the Office of Naval Research under grant N00014-21-1-2267,
and by the Army Research Office through the MIT Institute for Soldier Nanotechnologies, under Contract W911NF-13-D-0001.
Parts of this paper were previously presented at FUSION 2018, Cambridge, UK, July 2018 and at IEEE ICASSP 2019, Brighton, UK, May, 2019.
D.\ Gaglione, P.\ Braca, and G.\ Soldi are with the NATO Centre for Maritime Research and Experimentation (CMRE), La~Spezia, Italy (e-mail: domenico.gaglione@cmre.nato.int, paolo.braca@cmre.nato.int, giovanni.soldi@cmre.nato.int). 
F.\ Meyer is with the
Scripps Institution of Oceanography and the Department of Electrical and Computer Engineering, University of California San Diego, La Jolla, CA, USA (e-mail: flmeyer@ucsd.edu).
F.\ Hlawatsch is with the Institute of Telecommunications, TU Wien, Vienna, Austria (e-mail: franz.hlawatsch@tuwien.ac.at).
M.\ Z.\ Win is with the Laboratory for Information and Decision Systems (LIDS), Massachusetts Institute of Technology (MIT), Cambridge, MA, USA (e-mail: moewin@mit.edu).}
manuscript supplements the related manuscript, `Fusion of Sensor Measurements and Target-Provided Information in Multitarget Tracking'
by the same authors \cite{itself}.
The presented material comprises
in Section \ref{sec:TDRS_likelihood}
a recursive definition of the likelihood function $f(\V{q}_{n,m} | \V{x}_{n,k})$ involved in 
\cite[Eq.~(5)]{itself}
and
in Section \ref{sec:inner_loop_mex}
an efficient implementation of the iterative message passing algorithm for probabilistic data association
considered in
\cite[Sec.~V-D]{itself}.
Basic definitions, notation, and assumptions are given in \cite{itself} and will be repeated here only partly.\linebreak

\vspace{-5.5mm}

\section{Recursive Definition of $f(\V{q}_{n,m} | \V{x}_{n,k})$}
\label{sec:TDRS_likelihood}

In this section, we provide
a recursive definition of the target-dependent reporting system (TDRS) likelihood function $f(\V{q}_{n,m} | \V{x}_{n,k})$ 
involved in
\cite[Eq.~(5)]{itself}. Here, $\V{x}_{n,k}$ is the state of 
potential target (PT) $k$ at time $t_{n}$, and $\V{q}_{n,m} \! = \! \big[\V{q}^{(1)\T}_{n,m}, \ldots, \V{q}^{(L_{n,m})\T}_{n,m}\big]^{\T}$ is the vector of all the $L_{n,m}$ 
self-measurements in TDRS cluster $m$ at time $t_{n}$.
Each self-measurement $\V{q}_{n,m}^{(\ell)}$ is generated by a PT $k$ at some intermediate time $t_{n,m}^{(\ell)} \in ( t_{n-1},t_{n} ]$.
(We use the convention that $\ell < \ell'$ implies $t_{n,m}^{(\ell)} \leqslant t_{n,m}^{(\ell')}$.) 
We will also consider the state of PT $k$ at time $t_{n,m}^{(\ell)}$, denoted as $\V{x}^{(\ell)}_{n,k,m}$.
To simplify our notation in this section, we omit the time index $n$. Moreover, we consider one fixed PT and one fixed TDRS cluster and thus omit also the PT index $k$ 
and the cluster index $m$.
Accordingly, $f(\V{q}_{n,m} | \V{x}_{n,k})$, $\V{q}_{n,m} \! = \! \big[\V{q}^{(1)\T}_{n,m}, \ldots, \V{q}^{(L_{n,m})\T}_{n,m}\big]^{\T}\rmv$, $t_{n}$, $t_{n,m}^{(\ell)}$,
and $\V{x}^{(\ell)}_{n,k,m}$ are hereafter written as $f(\V{q} | \V{x})$, $\V{q} \rmv=\rmv \big[\V{q}^{(1)\T}\rmv, \ldots, \V{q}^{(L)\T}\big]^{\T}\rmv$, $t$, $t^{(\ell)}\rmv$, and $\V{x}^{(\ell)}\rmv$, respectively.

By the chain rule, the likelihood function $f(\V{q} | \V{x})$ can be factored 
as
\begin{align}
	f(\V{q} | \V{x}) = \prod_{\ell = 1}^{L} f\big(\V{q}^{(\ell)} \big| \V{q}^{(\ell + 1)}\rmv, \ldots, \V{q}^{(L)}\rmv, \V{x} \big).
	\label{eq:likelihood_AIS_kinematic_data} \\[-8mm]
	\nonumber
\end{align}
Each factor can be expressed by 
the law of total probability 
as 
\begin{align}
	f\big(\V{q}^{(\ell)} \big| \V{q}^{(\ell + 1)}\rmv, \ldots, \V{q}^{(L)}\rmv, \V{x} \big) 
	&=\rmv  \int \! f \big(\V{q}^{(\ell)} \big| \V{x}^{(\ell)}\rmv, \V{q}^{(\ell + 1)}, \ldots, \V{q}^{(L)}\rmv, \V{x}\big) \ist f \big(\V{x}^{(\ell)} \big| \V{q}^{(\ell + 1)}\rmv, \ldots, \V{q}^{(L)}\rmv, \V{x} \big) \ist d\V{x}^{(\ell)}.
	\label{eq:likelihood_AIS_kinematic_data_single_measurement_intermediate_step} \\[-8mm]
	\nonumber
\end{align}
Assuming that 
$\V{q}^{(\ell)}$ is conditionally independent, given $\V{x}^{(\ell)}\rmv$, of the PT state $\V{x}$ and of the subsequent self-measurements $\V{q}^{(\ell')}\rmv$, $\ell' \!>\rmv \ell$, 
we have $f \big(\V{q}^{(\ell)} \big| \V{x}^{(\ell)}\rmv, \V{q}^{(\ell + 1)}\rmv, \ldots, \V{q}^{(L)}\rmv, \V{x} \big) \! = \! f \big(\V{q}^{(\ell)} \big| \V{x}^{(\ell)}\big)$,
which is the self-measure\-ment likelihood function
introduced in
\cite[Sec.~III-B]{itself}.
Therefore, Eq.\ \eqref{eq:likelihood_AIS_kinematic_data_single_measurement_intermediate_step} 
becomes 
\begin{align}
	f\big(\V{q}^{(\ell)} \big| \V{q}^{(\ell + 1)}\rmv, \ldots, \V{q}^{(L)}\rmv, \V{x} \big) 
	=\rmv  \int \! f \big(\V{q}^{(\ell)} \big| \V{x}^{(\ell)}\big) \ist f \big(\V{x}^{(\ell)} \big| \V{q}^{(\ell + 1)}\rmv, \ldots, \V{q}^{(L)}\rmv, \V{x} \big) \ist d\V{x}^{(\ell)}.
	\label{eq:likelihood_AIS_kinematic_data_single_measurement} \\[-7.5mm]
	\nonumber
\end{align}
Here, the factor $f \big(\V{x}^{(\ell)} \big| \V{q}^{(\ell + 1)}\rmv, \ldots, \V{q}^{(L)}\rmv, \V{x}\big)$ can be developed by 
the law of total probability 
as
\begin{align}
\hspace{-6mm}	f \big( \V{x}^{(\ell)} \big| \V{q}^{(\ell + 1)}\rmv, \ldots, \V{q}^{(L)}\rmv, \V{x} \big) 
	& =\rmv \int \! f \big( \V{x}^{(\ell)} \big| \V{x}^{(\ell + 1)}\rmv, \V{q}^{(\ell + 1)}\rmv, \ldots, \V{q}^{(L)}\rmv, \V{x} \big) \ist f \big( \V{x}^{(\ell + 1)} \big| \V{q}^{(\ell + 1)}\rmv, \ldots, \V{q}^{(L)}\rmv, \V{x} \big) \ist d\V{x}^{(\ell + 1)}.
	\label{eq:likelihood_2nd_factor} \\[-7.5mm]
	\nonumber
\end{align}
Assuming that $\V{x}^{(\ell)}$ is conditionally independent, given $\V{x}^{(\ell + 1)}\rmv$, of $\V{x}$ and all $\V{q}^{(\ell')}\rmv$, $\ell' \!>\rmv \ell$,
we have $f \big( \V{x}^{(\ell)} \big| \V{x}^{(\ell + 1)}\rmv,$\linebreak $\V{q}^{(\ell + 1)}\rmv, \ldots, \V{q}^{(L)}\rmv, \V{x} \big) \! = \! f \big( \V{x}^{(\ell)} \big|  \V{x}^{(\ell + 1)} \big)$, which is the PT state transition probability density function (pdf) from 
$t^{(\ell + 1)}$ to $t^{(\ell)}\rmv$, corresponding to a ``backward-in-time'' dynamic model. Thus, Eq.\ \eqref{eq:likelihood_2nd_factor} 
becomes
\begin{align}
	f \big( \V{x}^{(\ell)} \big| \V{q}^{(\ell + 1)}\rmv, \ldots, \V{q}^{(L)}\rmv, \V{x} \big) 
	=\rmv \int \! f \big( \V{x}^{(\ell)} \big|  \V{x}^{(\ell + 1)} \big) \ist f \big( \V{x}^{(\ell + 1)} \big| \V{q}^{(\ell + 1)}\rmv, \ldots, \V{q}^{(L)}\rmv, \V{x} \big) \ist d\V{x}^{(\ell + 1)}.
	\label{eq:likelihood_2nd_factor_bis}\\[-7mm]
	\nonumber
\end{align}
Here, the factor $f \big( \V{x}^{(\ell + 1)} \big| \V{q}^{(\ell + 1)}\rmv, \ldots, \V{q}^{(L)}\rmv, \V{x} \big)$ can be developed using Bayes' rule
as
\begin{align}
	f \big( \V{x}^{(\ell + 1)} \big| \V{q}^{(\ell + 1)}\rmv, \ldots, \V{q}^{(L)}\rmv, \V{x} \big) 
&=\ist \frac{f \big( \V{q}^{(\ell + 1)} \big| \V{x}^{(\ell + 1)}\rmv, \V{q}^{(\ell + 2)}\rmv, \ldots, \V{q}^{(L)}\rmv, \V{x} \big) 
  \ist f \big( \V{x}^{(\ell + 1)} \big| \V{q}^{(\ell + 2)}\rmv, \ldots, \V{q}^{(L)}\rmv, \V{x} \big) 
  }{f \big( \V{q}^{(\ell + 1)} \big| \V{q}^{(\ell + 2)}\rmv, \ldots, \V{q}^{(L)}\rmv, \V{x} \big)} \nonumber \\[1.5mm]
	&=\ist \frac{f \big( \V{q}^{(\ell + 1)} \big| \V{x}^{(\ell + 1)}\big) \ist f \big( \V{x}^{(\ell + 1)} \big| \V{q}^{(\ell + 2)}\rmv, \ldots, \V{q}^{(L)}\rmv, \V{x} \big) 
	}{f \big( \V{q}^{(\ell + 1)} \big| \V{q}^{(\ell + 2)}\rmv, \ldots, \V{q}^{(L)}\rmv, \V{x} \big)} \ist,
	\label{eq:bayes_recursion} \\[-7mm]
	\nonumber	
\end{align}
where, in the last step,
our earlier assumption $f \big(\V{q}^{(\ell)} \big| \V{x}^{(\ell)}\rmv, \V{q}^{(\ell + 1)}\rmv, \ldots, \V{q}^{(L)}\rmv, \V{x} \big) \! = \! f \big(\V{q}^{(\ell)} \big| \V{x}^{(\ell)}\big)$ was
used.
Hence, Eq.\  \eqref{eq:likelihood_2nd_factor_bis} 
becomes
\be
\hspace{.6mm}f \big( \V{x}^{(\ell)} \big| \V{q}^{(\ell + 1)}\rmv, \ldots, \V{q}^{(L)}\rmv, \V{x} \big) 
	=\! \int \frac{f \big( \V{x}^{(\ell)} \big|  \V{x}^{(\ell + 1)} \big) \ist f \big( \V{q}^{(\ell + 1)} \big| \V{x}^{(\ell + 1)}\big) \ist f \big( \V{x}^{(\ell + 1)} \big| \V{q}^{(\ell + 2)}\rmv, \ldots, \V{q}^{(L)}\rmv, \V{x} \big) 
	}{f \big( \V{q}^{(\ell + 1)} \big| \V{q}^{(\ell + 2)}\rmv, \ldots, \V{q}^{(L)}\rmv, \V{x} \big)} \, d\V{x}^{(\ell + 1)}. \!\!\!\!\!
	\label{eq:likelihood_2nd_factor_1}
\ee

Eq.\  \eqref{eq:likelihood_2nd_factor_1} expresses $f \big( \V{x}^{(\ell)} \big| \V{q}^{(\ell + 1)}\rmv, \ldots, \V{q}^{(L)}\rmv, \V{x} \big)$ in terms of 
$f \big( \V{x}^{(\ell + 1)} \big| \V{q}^{(\ell + 2)}\rmv, \ldots, \V{q}^{(L)}\rmv, \V{x} \big)$.
Together with \eqref{eq:likelihood_AIS_kinematic_data_single_measurement}, this establishes a recursive procedure for obtaining 
$f\big(\V{q}^{(\ell)} \big| \V{q}^{(\ell + 1)}\rmv, \ldots, \V{q}^{(L)}\rmv, \V{x} \big)$ and, in turn, via \eqref{eq:likelihood_AIS_kinematic_data}, a definition of the likelihood function $f(\V{q} | \V{x})$. One recursion ($\ell \!+\! 1 \to \ell$, with $\ell \! \in \! \{ L-1, L-2, \ldots, 1 \}$) is given as follows:

\begin{itemize}

\item \emph{Input}: $f\big( \V{x}^{(\ell + 1)} \big| \V{q}^{(\ell + 2)}\rmv, \ldots, \V{q}^{(L)}\rmv, \V{x} \big)$ and $f\big( \V{q}^{(\ell + 1)} \big| \V{q}^{(\ell + 2)}\rmv, \ldots, \V{q}^{(L)}\rmv, \V{x} \big)$ 
calculated at the previous recursion;
self-measurement likelihood functions $f\big(\V{q}^{(\ell)} \big| \V{x}^{(\ell)}\big)$ and $f\big(\V{q}^{(\ell + 1)} \big| \V{x}^{(\ell + 1)}\big)$; PT state transition pdf
$f\big(\V{x}^{(\ell)} \big| \V{x}^{(\ell + 1)}\big)$.

\item \emph{Operations}:

\vspace{-1.5mm}

  \begin{enumerate}
  
  \item Obtain $f\big( \V{x}^{(\ell)} \big| \V{q}^{(\ell + 1)}\rmv, \ldots, \V{q}^{(L)}\rmv, \V{x} \big)$ according to \eqref{eq:likelihood_2nd_factor_1}.

	\item Obtain $f\big( \V{q}^{(\ell)} \big| \V{q}^{(\ell + 1)}\rmv, \ldots, \V{q}^{(L)}\rmv, \V{x} \big)$ according to \eqref{eq:likelihood_AIS_kinematic_data_single_measurement}.

  \end{enumerate}

\item \emph{Output}: $f\big( \V{x}^{(\ell)} \big| \V{q}^{(\ell + 1)}\rmv, \ldots, \V{q}^{(L)}\rmv, \V{x} \big)$ and $f\big( \V{q}^{(\ell)} \big| \V{q}^{(\ell + 1)}\rmv, \ldots, \V{q}^{(L)}\rmv, \V{x} \big)$.

\end{itemize}

\noindent This recursion is initialized at $\ell = L - 1$ with 
\[
f\big( \V{x}^{(\ell + 1)} \big| \V{q}^{(\ell + 2)}\rmv, \ldots, \V{q}^{(L)}\rmv, \V{x} \big) 
= f\big( \V{x}^{(L)} \big| \V{q}^{(L+ 1)}\rmv, \V{q}^{(L)}\rmv, \V{x} \big) = f \big( \V{x}^{(L)} \big| \V{x}\big)
\]
(note that in the condition, $(\V{q}^{(L+ 1)}\rmv, \V{q}^{(L)})$ is the empty list) and 
\begin{align*}
f\big( \V{q}^{(\ell + 1)} \big| \V{q}^{(\ell + 2)}\rmv, \ldots, \V{q}^{(L)}\rmv, \V{x} \big) &= f\big( \V{q}^{(L)} \big| \V{q}^{(L + 1)}\rmv, \V{q}^{(L)}\rmv, \V{x} \big) \\[1mm]
&= f \big( \V{q}^{(L)} \big| \V{x} \big)\\[.5mm]
&= \int \! f \big( \V{q}^{(L)} \big| \V{x}^{(L)}\rmv, \V{x}\big) \ist f \big( \V{x}^{(L)} \big| \V{x} \big) \ist d\V{x}^{(L)}\\[0mm]
&= \int \! f \big( \V{q}^{(L)} \big| \V{x}^{(L)}\big) \ist f \big( \V{x}^{(L)} \big| \V{x} \big) \ist d\V{x}^{(L)} ,
\end{align*}
where, in the last step, we used once again our earlier assumption $f \big(\V{q}^{(\ell)} \big| \V{x}^{(\ell)}\rmv, \V{q}^{(\ell + 1)}\rmv, \ldots, \V{q}^{(L)}\rmv, \V{x} \big) \! = \! f \big(\V{q}^{(\ell)} \big| \linebreak \V{x}^{(\ell)}\big)$ 
(with $\ell \!=\! L$). 
We note that $f \big( \V{x}^{(L)} \big| \V{x}\big)$ is the PT state transition pdf from $t$ to $t^{(L)}$
and $f \big( \V{q}^{(L)} \big| \V{x}^{(L)}\big)$ is the likelihood function of the $L$th self-measurement.

Thus, finally, $f(\V{q} | \V{x})$ is defined by recursively obtaining $f\big(\V{q}^{(\ell)} \big| \V{q}^{(\ell + 1)}\rmv, \ldots, \V{q}^{(L)}\rmv, \V{x} \big)$ for $\ell = L, L\!-\!1,\ldots, 1$ as stated above and then evaluating the product \eqref{eq:likelihood_AIS_kinematic_data}.

\section{Iterative Message Passing Algorithm for Probabilistic Data Association}
\label{sec:inner_loop_mex}

This section presents and derives an efficient implementation of the message calculation rules constituting the \linebreak ``iterative probabilistic data association'' part of the proposed 
multitarget tracking algorithm (cf.\
\cite[Sec.~V-D]{itself}).
We consider the inner loop of the factor graph shown in
\cite[Fig.~2]{itself}, which involves the variable nodes corresponding to the 
association variables $a_{n,k}^{(s)}$, $k \! \in \! \Set{K} \! = \! \{ 1,\ldots,K \}$ and $b_{n,m}^{(s)}$, $m \! \in \! \Set{M}_{n}^{(s)} \! = \! \big\{ 1,\ldots,M^{(s)}_{n} \big\}$.
As mentioned in
\cite[Sec.~V-D]{itself}, in the $j$th inner-loop iteration, messages $\xi_{k \rightarrow m}^{(s)(j)}\big(b_{n,m}^{(s)}\big)$ and 
$\nu_{m \rightarrow k}^{(s)(j)}\big(a_{n,k}^{(s)}\big)$ are calculated \linebreak for each PT $k \! \in \! \Set{K}$, data source $s \! \in \! \Set{S}_{0} \! = \! \{ 0,\ldots,S \}$, 
and observation $m \! \in \! \Set{M}_{n}^{(s)}$ according 
\vspace{-1mm}
to
\cite[Eq.~(25)]{itself}
\begin{align}
	\xi_{k \rightarrow m}^{(s)(j)}\big(b_{n,m}^{(s)}\big) &=\! \sum_{a_{n,k}^{(s)} = 0}^{M_{n}^{(s)}} \!\!\psi^{(s)}\big(a_{n,k}^{(s)},b_{n,m}^{(s)} \big) 
	\ist \beta^{(s)(i)} \big( a_{n,k}^{(s)} \big)
	\!\! \prod_{m' \in \Set{M}_{n}^{(s)} \setminus \{ m \}} \!\!\! \nu_{m' \rmv\rightarrow k}^{(s)(j-1)}\big(a_{n,k}^{(s)}\big)
	\label{eq:xi_full} \\[-11mm]
	\nonumber
\end{align}
and
\cite[Eq.~(26)]{itself}
\vspace{-3mm}
\begin{align}
	\nu_{m \rightarrow k}^{(s)(j)}\big(a_{n,k}^{(s)}\big) &=\! \sum_{b_{n,m}^{(s)} = 0}^{K} \!\!\psi^{(s)}\big(a_{n,k}^{(s)},b_{n,m}^{(s)} \big) 
    \!\rmv\prod_{k' \in \Set{K} \setminus \{ k \}} \!\!\! \xi_{k' \!\rightarrow m}^{(s)(j)}\big(b_{n,m}^{(s)}\big) \ist,
    \label{eq:nu_full} \\[-9mm]
    \nonumber
\end{align}
respectively.
Here, $i$ is the outer-loop iteration index.
The iteration constituted by \eqref{eq:xi_full} and \eqref{eq:nu_full} is initialized by $\nu_{m \rightarrow k}^{(s)(0)}\big(a_{n,k}^{(s)}\big) \! = \! 1$.
The efficient implementation of \eqref{eq:xi_full} and \eqref{eq:nu_full} to be presented in this section is a modification of the algorithm proposed in \cite{WilLau:J14}.

\subsection{A ``Binary'' Formulation}
\label{sec:binary}

Due to the definition of the binary function $\psi^{(s)}\big(a_{n,k}^{(s)},b_{n,m}^{(s)} \big)$
in
\cite[Sec.~III-C]{itself}, the expressions \eqref{eq:xi_full} and \eqref{eq:nu_full} can take on only two different values according to \cite{WilLau:J14}
\begin{align}
	\xi_{k \rightarrow m}^{(s)(j)}\big(b_{n,m}^{(s)}\big) =
	\begin{dcases}
		\xi_{k \rightarrow m,1}^{(s)(j)} \ist,	&	b_{n,m}^{(s)} \!\rmv=\! k \\[.3mm]
		\xi_{k \rightarrow m,2}^{(s)(j)} \ist,	&	b_{n,m}^{(s)} \!\rmv\neq\! k
	\end{dcases}
	\label{eq:xi_values} \\[-11mm]
	\nonumber
\end{align}
and
\vspace{-2mm}
\begin{align}
	\nu_{m \rightarrow k}^{(s)(j)}\big(a_{n,k}^{(s)}\big) =
	\begin{dcases}
		\nu_{m \rightarrow k,1}^{(s)(j)} \ist,	&	a_{n,k}^{(s)} \!\rmv=\! m \\[.3mm]
		\nu_{m \rightarrow k,2}^{(s)(j)} \ist,	&	a_{n,k}^{(s)} \!\rmv\neq\! m.
	\end{dcases}
	\label{eq:nu_values} \\[-8mm]
	\nonumber
\end{align}
Expressions of $\xi_{k \rightarrow m,1}^{(s)(j)}$ and $\xi_{k \rightarrow m,2}^{(s)(j)}$ are obtained by inserting the definition of the function $\psi^{(s)}\big(a_{n,k}^{(s)},b_{n,m}^{(s)} \big)$
and the expression \eqref{eq:nu_values} into \eqref{eq:xi_full}. One obtains after a straightforward 
calculation
\be
	\xi_{k \rightarrow m,1}^{(s)(j)} = \beta^{(s)(i)}_{n,k}(m) \! \prod_{m' \in \Set{M}_{n}^{(s)} \setminus \{ m \}} \!\!\! \nu_{m' \rmv\rightarrow k,2}^{(s)(j-1)}
	\label{eq:xi_ais_overline_1} 
\vspace{-5mm}
\ee
and
\be
	\xi_{k \rightarrow m,2}^{(s)(j)} = \beta^{(s)(i)}_{n,k}(0) \!\! \prod_{m' \in \Set{M}_{n}^{(s)} \setminus \{ m \}} \!\!\! \nu_{m' \rmv\rightarrow k,2}^{(s)(j-1)} 
	\ist+ \! \sum_{m' \in \Set{M}_{n}^{(s)} \setminus \{ m \}} \!\!\! \beta^{(s)(i)}_{n,k}(m') \ist \nu_{m' \rmv\rightarrow k,1}^{(s)(j-1)} \!\! \prod_{m'' \in \Set{M}_{n}^{(s)} \setminus \{ m,m' \}} \!\!\! \nu_{m'' \rmv\rightarrow k,2}^{(s)(j-1)} \ist,
	\label{eq:xi_ais_overline_2} 
\vspace{-1mm}
\ee
where the shorthand $\beta^{(s)(i)}_{n,k}(m) \rmv \deq \rmv \beta^{(s)(i)} (a^{(s)}_{n,k} \! = \! m)$ is 
used. Similarly, inserting the definition of $\psi^{(s)}\big(a_{n,k}^{(s)}, b_{n,m}^{(s)} \big)$ and the expression \eqref{eq:xi_values} into \eqref{eq:nu_full}, we 
\vspace{-1mm}
obtain
\be
	\nu_{m \rightarrow k,1}^{(s)(j)} = \! \prod_{k' \in \Set{K} \setminus \{ k \}} \!\!\rmv \xi_{k' \rmv\rightarrow m,2}^{(s)(j)}
	\label{eq:nu_ais_overline_1} 
\vspace{-5mm}
\ee
and
\vspace{2mm}
\be
	\nu_{m \rightarrow k,2}^{(s)(j)} = \mathds{1}[s \! \in \! \Set{S}]
		\!\rmv \prod_{k' \in \Set{K} \setminus \{ k \}} \!\!\rmv \xi_{k' \rmv\rightarrow m,2}^{(s)(j)} 
	\ist + \!\sum_{k' \in \Set{K} \setminus \{ k \}} \!\!\rmv \xi_{k' \rmv\rightarrow m,1}^{(s)(j)} \!\prod_{k'' \in \Set{K} \setminus \{ k,k' \}} \!\!\! \xi_{k'' \rmv\rightarrow m,2}^{(s)(j)} \ist.
	\label{eq:nu_ais_overline_2} 
\vspace{.5mm}
\ee
Here, the factor $\mathds{1}[s \! \in \! \Set{S}]$
is one for $s \! \in \! \Set{S} \! = \! \{ 1,\ldots,S \}$, i.e., for TIPS sensors, and zero for $s \! = \! 0$, i.e., for the TDRS.
This factor takes into account the no-false-alarms assumption valid for the TDRS (i.e., $b_{n,m}^{(0)} \!\neq\rmv 0$).

\subsection{An Efficient Message Passing Algorithm}
\label{sec:eff}

For a simplified 
calculation of \eqref{eq:xi_values} and \eqref{eq:nu_values}, we exploit the fact that messages can be normalized with no effect on the result \cite{Loe:04}.
Therefore,
we normalize $\xi_{k \rightarrow m}^{(s)(j)}\big(b_{n,m}^{(s)}\big)$ so that its value for $b^{(s)}_{n,m} \! \neq \! k$ is $1$, and
we normalize $\nu_{m \rightarrow k}^{(s)(j)}\big(a_{n,k}^{(s)}\big)$ so that its value for $a^{(s)}_{n,k} \! \neq \! m$ is $1$.
The normalized
messages are thus given by \cite{WilLau:J14}
\be
\bar{\xi}_{k \rightarrow m}^{(s)(j)}\big(b_{n,m}^{(s)}\big) \deq \frac{\xi_{k \rightarrow m}^{(s)(j)}\big(b_{n,m}^{(s)}\big) }{ \xi_{k \rightarrow m,2}^{(s)(j)} }
	\label{eq:xi_values_TIPS_bar_0} 
	\vspace{-2mm}
\ee
and
\vspace{2mm}
\be
\bar{\nu}_{m \rightarrow k}^{(s)(j)}\big(a_{n,k}^{(s)}\big) \deq \frac{\nu_{m \rightarrow k}^{(s)(j)}\big(a_{n,k}^{(s)}\big) }{ \nu_{m \rightarrow k,2}^{(s)(j)} } \ist.
	\label{eq:nu_values_TIPS_bar_0}
\vspace{-1mm}
\ee
Inserting \eqref{eq:xi_values} into \eqref{eq:xi_values_TIPS_bar_0}, it follows that
\be
	\bar{\xi}_{k \rightarrow m}^{(s)(j)}\big(b_{n,m}^{(s)}\big) =
	\begin{dcases}
		\bar{\xi}_{k \rightarrow m}^{(s)(j)} \ist,	&	b_{n,m}^{(s)} \!\rmv=\! k \\[.3mm]
		1 \ist,	&	b_{n,m}^{(s)} \!\rmv\neq\! k,
	\end{dcases}
\label{eq:xi_values_TIPS} 
\vspace{-1mm}
\ee
where 
\vspace{2mm}
\be
	\bar{\xi}_{k \rightarrow m}^{(s)(j)} \deq \frac{\xi_{k \rightarrow m,1}^{(s)(j)} }{ \xi_{k \rightarrow m,2}^{(s)(j)} }.
	\label{eq:xi_values_TIPS_bar} 
	\vspace{-1mm}
\ee
Similarly, inserting \eqref{eq:nu_values} into \eqref{eq:nu_values_TIPS_bar_0} results in
\be
	\bar{\nu}_{m \rightarrow k}^{(s)(j)}\big(a_{n,k}^{(s)}\big) =
	\begin{dcases}
		\bar{\nu}_{m \rightarrow k}^{(s)(j)} \ist,	&	a_{n,k}^{(s)} \!\rmv=\! m \\[0mm]
		1 \ist,	&	a_{n,k}^{(s)} \!\rmv\neq\! m,
	\end{dcases}
	\label{eq:nu_values_TIPS}
\pagebreak \ee
where 
\be
\bar{\nu}_{m \rightarrow k}^{(s)(j)} \deq \frac{\nu_{m \rightarrow k,1}^{(s)(j)} }{ \nu_{m \rightarrow k,2}^{(s)(j)} } \ist.
	\label{eq:nu_values_TIPS_bar}
\ee
Finally, inserting \eqref{eq:xi_ais_overline_1} and \eqref{eq:xi_ais_overline_2}
into \eqref{eq:xi_values_TIPS_bar} and using the normalized message $\bar{\nu}_{m \rmv\rightarrow k}^{(s)(j)}(a^{(s)}_{n,k})$ (cf. \eqref{eq:nu_values_TIPS})
\vspace{-2mm}
gives
\be
	\bar{\xi}_{k \rightarrow m}^{(s)(j)} = \frac{ \beta^{(s)(i)}_{n,k}(m)}{\beta^{(s)(i)}_{n,k}(0) + \sum_{m' \in \Set{M}_{n}^{(s)} \setminus \{ m \}} \rmv\beta^{(s)(i)}_{n,k}(m') \ist \bar{\nu}_{m' \rmv\rightarrow k}^{(s)(j-1)}} \ist.
	\label{eq:xi_bar} 
\vspace{-1mm}
\ee
Similarly, inserting \eqref{eq:nu_ais_overline_1} and \eqref{eq:nu_ais_overline_2}
into \eqref{eq:nu_values_TIPS_bar} and using the normalized message $\bar{\xi}_{k \rmv\rightarrow m}^{(s)(j)}(b^{(s)}_{n,m})$ (cf. \eqref{eq:xi_values_TIPS})
\vspace{-1mm}
gives
\be
	\bar{\nu}_{m \rightarrow k}^{(s)(j)} 
	= \frac{1}{\mathds{1}[s \! \in \! \Set{S}]
		+ \sum_{k' \in \Set{K} \setminus \{ k \}}\rmv \bar{\xi}_{k' \rmv\rightarrow m}^{(s)(j)}} \ist.
	\label{eq:nu_bar} 
\vspace{-1mm}
\ee
The iteration
established by \eqref{eq:xi_bar} and \eqref{eq:nu_bar} is initialized by $\bar{\nu}_{m\rightarrow k}^{(s)(0)} \! = \! 1$.
This iteration
constitutes the efficient implementation of \eqref{eq:xi_full} and \eqref{eq:nu_full} for $s \! \in \! \Set{S}$, i.e., for the TIPS sensors; however, as we will explain next,
a modified iteration has to be used for the TDRS.

\subsection{Modification for $s \! = \! 0$}
\label{sec:eff_0}

To within a normalization, $\beta^{(s)(i)} \big( a_{n,k}^{(s)} \big)$ can be interpreted as a
single-PT association probability prior to data association \cite{williams10}. Thus, we can assume that $\beta^{(s)(i)}_{n,k}(0) \rmv = \rmv \beta^{(s)(i)} \big( a_{n,k}^{(s)} \!\rmv = \! 0 \big)$
is nonzero; for $s \! \in \! \Set{S}$, this expresses the fact 
that there is a nonzero probability that PT $k$ is not detected by TIPS sensor $s$, and for $s \! = \! 0$, it means that there is a nonzero probability that no TDRS report is generated by PT $k$.
As a consequence,
the denominator of expression \eqref{eq:xi_bar} is nonzero. 
The denominator of expression \eqref{eq:nu_bar}, on the other hand, is nonzero for $s \! \in \! \Set{S}$ but it may be zero for $s \! = \! 0$.
For this reason, the iterative algorithm given by \eqref{eq:xi_bar} and \eqref{eq:nu_bar}---which is analogous to that proposed in \cite{WilLau:J14}---can be used for $s \! \in \! \Set{S}$, 
i.e., for the TIPS sensors, but not for $s \! = \! 0$, i.e., for the TDRS.

Therefore, for $s \! = \! 0$, we propose a different iterative algorithm that is based on
an alternative normalization of
the message $\nu^{(0)(j)}_{m \rightarrow k} (a^{(0)}_{n,k})$.
We now normalize $\nu^{(0)(j)}_{m \rightarrow k} (a^{(0)}_{n,k})$
such that its value for $a^{(0)}_{n,k} \! = \! m$ is $1$, 
which yields
\vspace{-1mm}
\be
\breve{\nu}_{m \rightarrow k}^{(0)(j)}\big(a_{n,k}^{(0)}\big) \deq \frac{\nu_{m \rightarrow k}^{(0)(j)}\big(a_{n,k}^{(0)}\big) }{ \nu_{m \rightarrow k,1}^{(0)(j)} } \ist.
	\label{eq:nu_values_TIPS_bar_0_TDRS}
\vspace{-2mm}
\ee
Inserting \eqref{eq:nu_values} into  \eqref{eq:nu_values_TIPS_bar_0_TDRS},
it follows that
\vspace{-1mm}
\be
	\breve{\nu}_{m \rightarrow k}^{(0)(j)}\big(a_{n,k}^{(0)}\big) =
	\begin{dcases}
		1 \ist,	&	a_{n,k}^{(0)} \!\rmv=\! m \\[0mm]
		\breve{\nu}_{m \rightarrow k}^{(0)(j)} \ist,	&	a_{n,k}^{(0)} \!\rmv\neq\! m,
	\end{dcases}
	\label{eq:nu_values_TDRS} 
\vspace{-2mm}
\ee
where
\vspace{2mm}
\be
\breve{\nu}_{m \rightarrow k}^{(0)(j)} \deq \frac{ \nu_{m \rightarrow k,2}^{(0)(j)} }{ \nu_{m \rightarrow k,1}^{(0)(j)} } 
\ist.
	\label{eq:nu_values_TDRS_bar}
\vspace{.5mm}
\ee
Comparing \eqref{eq:nu_values_TDRS_bar} with \eqref{eq:nu_values_TIPS_bar} and, in turn, using \eqref{eq:nu_bar} and the fact that $\mathds{1}[0 \! \in \! \Set{S}] \! = \! 0$, we obtain
\vspace{1mm}
\be
\breve{\nu}_{m \rightarrow k}^{(0)(j)} = \frac{1}{\bar{\nu}^{(0)(j)}_{m \rightarrow k}} =\! \sum_{k' \in \Set{K} \setminus \{ k \}}\rmv\!\! \bar{\xi}_{k' \rmv\rightarrow m}^{(0)(j)} \ist.
	\label{eq:nu_values_TDRS_bar_1}
\ee
Furthermore, using the relation $\bar{\nu}^{(0)(j)}_{m \rightarrow k} = 1/\breve{\nu}_{m \rightarrow k}^{(0)(j)}$ (see
\eqref{eq:nu_values_TDRS_bar_1}), 
expression \eqref{eq:xi_bar} for $s \!=\! 0$ can be rewritten in terms of $\breve{\nu}_{m \rightarrow k}^{(0)(j-1)}$
\vspace{-3mm}
as
\be
	\bar{\xi}_{k \rightarrow m}^{(0)(j)} = \frac{ \beta^{(0)(i)}_{n,k}(m)}{\beta^{(0)(i)}_{n,k}(0) + \sum_{m' \in \Set{M}_{n}^{(0)} \setminus \{ m \}} \rmv\beta^{(0)(i)}_{n,k}(m') / \breve{\nu}_{m' \rmv\rightarrow k}^{(0)(j-1)}} \ist.
	\label{eq:xi_breve} 
\vspace{-1mm}
\ee
Since, as before,
$\beta^{(0)(i)}_{n,k}(0)$ is nonzero, the denominator of \eqref{eq:xi_breve} is nonzero.
The iteration
established by \eqref{eq:nu_values_TDRS_bar_1} and \eqref{eq:xi_breve} is initialized by $\breve{\nu}_{m\rightarrow k}^{(0)(0)} \! = \! 1$.
This iteration
constitutes the efficient implementation of \eqref{eq:xi_full} and \eqref{eq:nu_full} for $s \! = \! 0$, i.e., for the TDRS.

An efficient MATLAB implementation of \eqref{eq:xi_bar} and \eqref{eq:nu_bar} for $s \! \in \! \Set{S}$ was developed in \cite{williams10}. This implementation 
has a complexity of order $O(K M_{n}^{(s)})$ per inner-loop iteration $j$
and avoids
the direct calculation of 
$\bar{\xi}_{k \rightarrow m}^{(s)(j)}$ 
and
$\bar{\nu}_{m \rightarrow k}^{(s)(j)}$ 
for all $m \! \in \! \Set{M}_{n}^{(s)}$ and all $k \! \in \! \Set{K}$. 
A similar implementation of \eqref{eq:nu_values_TDRS_bar_1} and \eqref{eq:xi_breve} with a complexity of order $O(K M_{n}^{(0)})$ can be obtained through simple modifications.

\subsection{Calculation of the Messages $\eta^{(s)(i)}\big(a_{n,k}^{(s)}\big)$}
\label{sec:eta}

After all the inner-loop message passing iterations $j = 1,\ldots,J$ have been carried out, the messages $\bar{\nu}_{m \rightarrow k}^{(s)(J)}$, $s \! \in \! \Set{S}$
and $\breve{\nu}_{m \rightarrow k}^{(0)(J)}$ are available. Next, for all data sources $s \! \in \! \Set{S}_0$, messages $\eta^{(s)(i)}\big(a_{n,k}^{(s)}\big)$, $k \!\in\! \Set{K}$ are calculated as 
\cite[Eq.~(27)]{itself}
\be
	\eta^{(s)(i)}\big(a_{n,k}^{(s)}\big) =\! \prod_{m \in \Set{M}_{n}^{(s)}} \!\!\rmv \nu_{m \rightarrow k}^{(s)(J)}\big(a_{n,k}^{(s)}\big) \ist.
	\label{eq:eta} 
\vspace{1mm}
\ee

Consider a TIPS sensor $s \! \in \! \Set{S}$. Using \eqref{eq:nu_values_TIPS_bar_0}, we can rewrite \eqref{eq:eta} in terms of the normalized messages 
$\bar{\nu}_{m \rightarrow k}^{(s)(J)}\big(a_{n,k}^{(s)}\big)$, 
\vspace{-1.5mm}
i.e.,
\be
	\eta^{(s)(i)}\big(a_{n,k}^{(s)}\big) =\! \prod_{m \in \Set{M}_{n}^{(s)}} \!\!\rmv \bar{\nu}_{m \rightarrow k}^{(s)(J)}\big(a_{n,k}^{(s)}\big) \ist \nu_{m \rightarrow k,2}^{(s)(J)} 
	\ist\propto \!\prod_{m \in \Set{M}_{n}^{(s)}} \!\!\rmv \bar{\nu}_{m \rightarrow k}^{(s)(J)}\big(a_{n,k}^{(s)}\big) \ist,
	\label{eq:eta_normalized_TIPS} 
\vspace{-1mm}
\ee
where we suppressed the constant factor $\prod_{m \in \Set{M}_{n}^{(s)}} \nu_{m \rightarrow k,2}^{(s)(J)}$.
Then, recalling from \eqref{eq:nu_values_TIPS} that
$\bar{\nu}_{m \rightarrow k}^{(s)(J)}\big(a_{n,k}^{(s)}\big) \rmv= \bar{\nu}_{m \rightarrow k}^{(s)(J)}$ if $a^{(s)}_{n,k} \! = \! m$ and $\bar{\nu}_{m \rightarrow k}^{(s)(J)}\big(a_{n,k}^{(s)}\big) \rmv=\rmv 1$ 
if $a^{(s)}_{n,k} \! \neq \! m$, we obtain for $a^{(s)}_{n,k} \! = \! m \! \in \Set{M}^{(s)}_{n}$
\[
\eta^{(s)(i)} \big( a^{(s)}_{n,k} \! = \! m \big) \propto\! \prod_{m' \in \Set{M}_{n}^{(s)}} \!\!\rmv \bar{\nu}_{m' \rightarrow k}^{(s)(J)} \big( a^{(s)}_{n,k} \! = \! m \big) 
  =\ist \bar{\nu}_{m \rightarrow k}^{(s)(J)} \big( a^{(s)}_{n,k} \! = \! m \big) \!\!\prod_{m' \in \Set{M}_{n}^{(s)} \setminus \{ m \}} \!\!\! \bar{\nu}_{m' \rightarrow k}^{(s)(J)} \big( a^{(s)}_{n,k} \! = \! m \big) 
  =\ist \bar{\nu}_{m \rightarrow k}^{(s)(J)} ,
 \vspace{-3mm}
 \]
and for $a^{(s)}_{n,k} \! = \! 0$
\[
\eta^{(s)(i)} \big( a^{(s)}_{n,k} \! = \! 0 \big) \propto\! \prod_{m' \in \Set{M}_{n}^{(s)}} \!\!\rmv \bar{\nu}_{m' \rightarrow k}^{(s)(J)} \big( a^{(s)}_{n,k} \! = \! 0 \big) 
  = 1 \ist.
\]
Thus, for $s \! \in \! \Set{S}$, the messages $\eta^{(s)(i)} \big( a^{(s)}_{n,k} \big)$ are given by
\begin{align}
	\eta^{(s)(i)} \big( a^{(s)}_{n,k} \big) \propto
	\begin{dcases}
		\bar{\nu}_{m \rightarrow k}^{(s)(J)} \ist, &  a_{n,k}^{(s)} \!=\rmv m \rmv\in\! \Set{M}_{n}^{(s)} \\[0mm]
		1, &  a_{n,k}^{(s)} \!=\rmv 0\ist.
	\end{dcases}
\end{align}

A similar expression is obtained for the TDRS ($s \! = \! 0$). Using \eqref{eq:nu_values_TIPS_bar_0_TDRS}, we can rewrite \eqref{eq:eta} in terms of the normalized messages 
$\breve{\nu}_{m \rightarrow k}^{(0)(J)} \big( a^{(0)}_{n,k} \big)$, i.e.,
\begin{align}
	\eta^{(0)(i)}\big(a_{n,k}^{(0)}\big) =\! \prod_{m \in \Set{M}_{n}^{(0)}} \!\!\rmv \breve{\nu}_{m \rightarrow k}^{(0)(J)}\big(a_{n,k}^{(0)}\big) \ist \nu_{m \rightarrow k,1}^{(0)(J)}
	\ist\propto \!\prod_{m \in \Set{M}_{n}^{(0)}} \!\!\rmv \breve{\nu}_{m \rightarrow k}^{(0)(J)}\big(a_{n,k}^{(0)}\big) \ist.
	\label{eq:eta_normalized_TDRS} 
\end{align}
Recalling from \eqref{eq:nu_values_TDRS} that
$\breve{\nu}_{m \rightarrow k}^{(0)(J)}\big(a_{n,k}^{(0)}\big) \rmv=\rmv 1$ if $a^{(0)}_{n,k} \! = \! m$ and 
$\breve{\nu}_{m \rightarrow k}^{(0)(J)}\big(a_{n,k}^{(0)}\big) \rmv=\rmv \breve{\nu}_{m \rightarrow k}^{(0)(J)}$ if $a^{(0)}_{n,k} \! \neq \! m$,
we obtain for $a^{(0)}_{n,k} \! = \! m \! \in \Set{M}^{(0)}_{n}$
\vspace{-2mm}
\begin{align*}
	\eta^{(0)(i)} \big( a^{(0)}_{n,k} \! = \! m \big) &\propto\! \prod_{m' \in \Set{M}_{n}^{(0)}} \!\!\rmv \breve{\nu}_{m' \rightarrow k}^{(0)(J)} \big( a^{(0)}_{n,k} \! = \! m \big) \\[1.5mm]
	&=\ist \breve{\nu}_{m \rightarrow k}^{(0)(J)} \big( a^{(0)}_{n,k} \! = \! m \big) \!\!\prod_{m' \in \Set{M}_{n}^{(0)} \setminus \{ m \}} 
	   \!\!\! \breve{\nu}_{m' \rightarrow k}^{(0)(J)} \big( a^{(0)}_{n,k} \! = \! m \big)\\[1mm]
	&= \prod_{m' \in \Set{M}_{n}^{(0)} \setminus \{ m \}} \!\!\! \breve{\nu}_{m' \rightarrow k}^{(0)(J)} \ist, \\[-10mm]
 \end{align*}
 and for 
\vspace{1mm}
$a^{(0)}_{n,k} \! = \! 0$
\[
	\eta^{(0)(i)} \big( a^{(0)}_{n,k} \! = \! 0 \big) \propto\! \prod_{m' \in \Set{M}_{n}^{(0)}} \!\!\rmv \breve{\nu}_{m' \rightarrow k}^{(0)(J)} \big( a^{(0)}_{n,k} \! = \! 0 \big) 
	=\! \prod_{m' \in \Set{M}_{n}^{(0)}} \!\!\rmv \breve{\nu}_{m' \rightarrow k}^{(0)(J)} \ist .
\]
That is,
\begin{align}
	\eta^{(0)(i)} \big( a^{(0)}_{n,k} \big) \propto
	\begin{dcases}
		\prod_{m' \in \Set{M}_{n}^{(0)} \setminus \{ m \}} \!\!\! \breve{\nu}_{m' \rightarrow k}^{(0)(J)}, &  a_{n,k}^{(0)} \!=\rmv m \rmv\in\! \Set{M}_{n}^{(0)} \\[2mm]
		\prod_{m' \in \Set{M}_{n}^{(0)}} \!\!\rmv \breve{\nu}_{m' \rightarrow k}^{(0)(J)}, &  a_{n,k}^{(0)} \!=\rmv 0\ist.
	\end{dcases}
\end{align}

\bibliographystyle{IEEEtran}
\bibliography{IEEEabrv,biblio}